\documentclass[mathematics,article,submit,pdftex,moreauthors]{Definitions/mdpi} 
\renewcommand{\linenumbers}{}
\firstpage{1} 
\pubvolume{1}
\issuenum{1}
\articlenumber{0}
\pubyear{2022}
\copyrightyear{2022}
\datereceived{} 
\dateaccepted{} 
\datepublished{} 
\hreflink{https://doi.org/} % If needed use \linebreak
\Title{Numerical study of a confined vesicle in shear flow 
at finite temperature}

\TitleCitation{Numerical study of a confined vesicle in shear flow 
at finite temperature}

\Author{Antonio Lamura $^{1}$\orcidA{}}

\AuthorNames{Antonio Lamura}

\AuthorCitation{Lamura, A.}
\address{%
$^{1}$ \quad Istituto Applicazioni Calcolo, CNR, Via Amendola 122/D, 70126 Bari, Italy; antonio.lamura@cnr.it}

\abstract{The dynamics and rheology of a vesicle confined in a channel under shear flow
are studied at finite temperature. The effect of finite 
temperature on vesicle motion and system viscosity is investigated.
A two-dimensional numerical model, which includes thermal fluctuations and is 
based on a combination of molecular dynamics
and mesoscopic hydrodynamics, is used to perform a detailed analysis
in a wide range of the Peclet numbers (the ratio of the shear rate 
to the rotational diffusion coefficient). 
The suspension viscosity
is found to be a monotonous increasing function of the viscosity contrast 
(the ratio of the viscosity of the encapsulated fluid to that of 
the surrounding fluid) both in the tank-treading and the tumbling regime
due to the interplay of different temperature-depending mechanisms.
Thermal effects induce shape and inclination fluctuations of the vesicle which experiences also
Brownian diffusion across the channel increasing the viscosity.
These effects reduces when increasing the Peclet number.
}


\begin{document}

%%%%%%%%%%%%%%%%%%%%%%%%%%%%%%%%%%%%%%%%%%

\section{Introduction}

Suspensions of soft particles such as droplets, vesicles, and capsules 
are ubiquitous in
relevant applications in biology, medicine, and engineering.
Studying their dynamics in flow is challenging, since
shapes are not fixed, as in the case of rigid objects, 
but depend dynamically on the interplay between fluid stresses and interfacial
forces. The interfacial forces are directly related to the 
nature of the considered particles: 
The surface tension for droplets, the membrane bending
rigidity for vesicles, and additionally the membrane shear elasticity for
capsules. This calls for separate investigations of the various 
systems. 

Vesicles are small volumes of fluid embedded in a lipid bi-layer
membrane, in solution with either the same or different fluid.
The dynamical and rheological properties of their suspensions 
in flow have attracted a lot of theoretical and experimental
interest, as 
comprehensively reviewed in 
Refs.~\cite{vlah09,abre14,wink14,bies16}.
A consensus has been reached concerning the dynamical regimes
in shear flow. In dilute solution, vesicles can
show tank-treading (TT), tumbling (TU), and vacillating-breathing (VB)
(also called trembling or swinging) motion, depending on the shear rate
and the viscosity contrast $\lambda=\eta_{out}/\eta_{in}$,
where $\eta_{in}$ and $\eta_{out}$ are the viscosities 
of the inner and outer fluids, respectively. 
TT and TU occur at low and high $\lambda$, respectively,
while VB appears for strong flows when vesicle deformation
affects its dynamics \cite{kell82,nogu04,nogu05,kant05,kant06,
misb06,nogu07bis,lebe07,vlah07,mess09,zhao11}.

On the other hand, 
the rheology of single vesicle suspensions is still a matter of debate. 
Indeed,
different behaviors of the intrinsic viscosity 
$\eta_I=(\eta - \eta_{out})/(\eta_{out} \phi)$, where
$\eta$ is the effective system viscosity and $\phi$
the vesicle concentration, 
as a function of the viscosity
contrast have been observed in experimental, theoretical, 
and numerical studies. 
In the case of very dilute suspensions of quasi-spherical vesicles,
it was shown analytically \cite{dank07a,dank07b}
that the intrinsic viscosity decreases with the viscosity contrast $\lambda$
in the TT regime, reaching a minimum at the TT-to-TU transition,
and then grows with $\lambda$ in the TU regime.
Experimental investigations do not provide conclusive results. A good agreement
with the theoretical prediction  was found in Ref.~\cite{vitk08},
while an increase of $\eta_I$ with $\lambda$ for $\lambda < 1$ was observed
in Ref.~\cite{kant08}. These discrepancies might be due to the difficulty
in preparing monodisperse suspensions as well as to the fact that
viscosity measurements require volume fractions $\phi \sim 5\% - 10\%$, thus
making the extrapolation to the dilute limit difficult \cite{kant08}.
Numerical models differ mainly in the lack or presence of thermal noise.
In the former case, it was found in two-dimensional models that the
intrinsic viscosity follows the theoretical prediction 
both in the very dilute \cite{ghig10,kaou14,nait19} and in the dilute
case \cite{rahi10,thie13,kaou14}. A similar dependence on the viscosity contrast
was found also in a three-dimensional model \cite{zhao13}.
The only available numerical model with thermal fluctuations
\cite{lamu13} shows that $\eta_I$ is an increasing function of $\lambda$,
in agreement with the experiments of Ref.~\cite{kant08}.

The numerical model of Ref.~\cite{lamu13}, 
which comprises both thermal membrane 
undulations and thermal noise \cite{lamu13}, 
is adopted here to perform a detailed study of a confined vesicle 
in shear flow
at finite temperature. The results of
this model yielded very good agreement with experimental results
in describing the collision process of two vesicles  \cite{kant08}
and the flow field of a single vesicle
in shear flow \cite{afik16}.
The system is studied in two dimensions
at fixed shear rate in a wide range of the Peclet number
$Pe$ - 
the ratio of the shear rate to the rotational diffusion coefficient -, 
differently
from other theoretical and numerical studies where $Pe=\infty$.
We aim at elucidating the role
played by thermal fluctuations in influencing both the vesicle dynamics and,
consequently, 
the system viscosity in the TT and TU regimes. 
The reason of considering a very dilute solution is
twofold. On one hand this allows the matching with the hypothesis
 of extremely dilute suspension used in the theoretical model
\cite{dank07a,dank07b}, and, on the other hand,
hydrodynamic and steric interactions between vesicles can be ruled out.

The paper is organized as follows. Section 2 presents the numerical model.
Results are illustrated in Section 3. A detailed discussion of our findings
about the effects of thermal noise is presented in Section 4, including
a comparison with previous studies. Finally, conclusions are presented in Section 5.

%%%%%%%%%%%%%%%%%%%%%%%%%%%%%%%%%%%%%%%%%%
\section{The model}

A two-dimensional fluid made of
$N_s$ point-like particles of mass $m$ is considered.
The particle  positions ${\bf r}_i(t)$ and
velocities ${\bf v}_i(t)$, $i=1,2,...,N_s$, at time $t$ 
are continuous variables. We employ the multi-particle collisions (MPC)
dynamics approach, in which
the time evolution occurs via iterative propagations and collisions
\cite{male99,male00,kapr08,gomp09}.
In the first streaming step, particles are ballistically
streamed for a time interval $\Delta t_s$
\begin{equation}
{\bf r}_i(t+\Delta t_s)
={\bf r}_i(t)+{\bf v}_i(t) \Delta t_s \;\;\;\;\; i=1,...,N_s .
\label{eq.prop}
\end{equation}
In the subsequent collision step, the system is divided into square cells 
of mesh size $a$ where an
instantaneous multi-particle collision occurs, which changes
particle velocities as
\begin{equation}
{\bf v}_i^{new}={\bf v}_c^G + {\bf v}_i^{ran} 
- \sum_{j \in cell} {\bf v}_j^{ran} / N_c 
+ {\bf \Pi}^{-1} \sum_{j \in cell} m \left [ {\bf r}_{j,c} \times 
({\bf v}_j - {\bf v}_j^{ran}) \right ] \times {\bf r}_{i,c} \;\;\;\;\; i=1,...,N_s
\label{eq.coll}
\end{equation}
where ${\bf v}_c^G$ is the center-of-mass velocity of all particles 
in the cell, ${\bf v}_i^{ran}$ is a velocity taken from a Maxwell-Boltzmann
distribution, $N_c$ is the number of particles in the cell, ${\bf \Pi}$
and ${\bf r}_{i,c}$ are the
moment-of-inertia tensor and 
the position relative to the center of mass of the particles in the cell,
respectively. 
This dynamics conserves
both local linear and angular momentum \cite{nogu07,goetze07} and keeps
the temperature constant \cite{alla02}.
The viscosity of the fluid is given by \cite{nogu08}
\begin{equation}
\eta=\frac{m}{\Delta t_s}
\Big [ \Big ( \frac{l}{a} \Big )^2 \Big ( \frac{n^2}{n-1} 
-\frac{n}{2}\Big ) 
+ \frac{1}{24} \Big (n - \frac{7}{5} \Big ) \Big ]
\label{visc}
\end{equation}
$n$ being the average number of particles per cell, 
$l=\Delta t_s \sqrt{k_B T /m}$ the mean-free path, 
and $k_B T$ the thermal energy.
The system of size $L_x \times L_y$ is confined
between two
horizontal walls sliding
along the $x$ direction
with velocities $v_{wall}$ and $-v_{wall}$.
Periodic boundary conditions (BC)
are used along the $x$ direction. 
Bounce-back BC are enforced at walls \cite{lamu01}
obtaining a linear flow profile $(u_x, u_y)=(\dot\gamma y, 0)$ with shear
rate $\dot\gamma= 2 v_{wall} / L_y$.

The vesicle membrane 
is modeled as a chain of $N_p$ beads of mass $m_p$ connected
to form a closed ring with average bond length $r_0$.
Neighboring beads interact via an harmonic potential
\begin{equation}
U_{bond}=\kappa_h \sum_{i=1}^{N_p} 
\frac{(|{\bf r}_i-{\bf r}_{i-1}|-r_0)^2}{2 r_0^2} 
\end{equation}
where $\kappa_h$ is the spring constant and ${\bf r}_i$ is the position vector
of the $i$-th bead. This ensures the conservation of the membrane length.
Shapes and fluctutions are controlled by the bending potential
\begin{equation}
U_{bend}=\frac{\kappa}{r_0} \sum_{i=1}^{N_p} (1-\cos \beta_{i}) ,
\label{bend}
\end{equation}
where $\kappa$ is the bending rigidity and $\beta_{i}$ 
is the angle between two consecutive bonds.
Finally, the internal area $S$
is kept close
to the target area $S_0$ of the
vesicle by using a quadratic constraint-potential with compression modulus
$\kappa_S$ \cite{lamu13}
\begin{equation}
U_{area} = \kappa_S \frac{(S-S_0)^2}{2 r_0^4} .
\label{area_pot}
\end{equation}
Newton's equations of motions of beads are integrated by using
the velocity-Verlet algorithm with time step $\Delta t_p$ \cite{allen}. 

In order to describe the coupling of solvent particles with the vesicle,
each bead is treated as a ``rough" hard disk having radius $r_v$ 
\cite{fink08,lamu13,lamu15}.
The value of $r_v$ is set so that disks overlap obtaining
a full covering up of the membrane. 
Scattering takes place when a solvent particle $i$ and a disk $j$
overlap while moving towards each other so that both the conditions 
$|{\bf r}_j-{\bf r}_i| < r_v$ and 
$ ({\bf r}_j-{\bf r}_i) \cdot ({\bf v}_j-{\bf v}_i) < 0$ are fulfilled. 
A second disk $k=j \pm 1$, connected to the $j$-th one and characterized
by the smallest distance from the solvent particle $i$, is then selected.
The angular velocity
\begin{equation}
{\bf \Omega} 
= {\bf \Pi}^{-1} \sum_{l=i,j,k} m_l {\bf r}_{l,c} \times {\bf v}_l
\end{equation}
and the center of mass velocity ${\bf v}^{G}$
of the $i,j,k$-particle system are computed,
${\bf r}_{l,c}$ being the position relative to the
center of mass.  The updated values of the velocities are given by
\begin{equation}
{\bf v}_l^{new} = 2 ({\bf v}^{G} + {\bf \Omega} \times {\bf r}_{l,c})
- {\bf v}_l \;\;\;\;\; l=i,j,k
\label{eq.mix}
\end{equation} 
which guarantees linear and angular momenta conservation \cite{mess09}.
The collision step (\ref{eq.coll}) is then performed for those fluid
particles which did not interact with the membrane in order to avoid 
multiple collisions with the same membrane disk in the following iterations.
Disks interact with lateral walls also by implementing the bounce-back scattering. The numerical implementation of the algorithm is outlined in Appendix A.

Inertial effects, which are experimentally irrelevant due to the small
flow velocities, are made negligible in the simulations
by making the Reynolds number 
$Re=\dot\gamma \rho R_0^2 / \eta_{out}$, with mass density
$\rho$, very small.
Other relevant dimensionless quantities 
are the reduced area $S^*=S_0/\pi R_0^2$, where
$R_0=L_0/2 \pi$ is the vesicle radius with $L_0$ the vesicle contour
length, and the reduced
shear rate $\dot \gamma^*=\dot\gamma \tau_c$, where $\tau_c=\eta_{out} R_0^3/\kappa$
is the relaxation time of the vesicle.
The viscosity contrast can be approximated as
$\lambda \simeq m_{in}/m_{out}$ within the present model
\cite{nogu07} (the subscripts
$out/in$ will refer to quantities outside/inside vesicle).
We use in the following 
$L_x=18.95 R_0$, $L_y=5.79 R_0$ with $R_0=7.6 a$.
Finally, we set 
$m_{in}$ such as to obtain $0.1 \leq \lambda \leq 15.0$, $m_p=3 m_{out}$, 
$\Delta t_s/\Delta t_p=64$, $N_p=480$, $r_v=r_0=a/10$,
$\kappa_S= 4 \times 10^{-4} k_B T$, $\kappa_h=3 \times 10^2 k_B T$. 
The setting of parameters is such to have
$Re < 0.15$, Mach number $Ma=v_{wall}/c_s < 0.25$, where 
$c_s=\sqrt{2 k_B T/m_{out}}$
is the speed of sound, to reduce compressibility effects
\cite{lamu02}, 
and $\dot\gamma^*=1.0$ in all the cases.
The value of the reduced shear rate  $\dot\gamma^*$ is comparable to those 
used in other studies \cite{ghig10,zhao13,thie13,kaou14} and allows 
the access of the TT and TU regimes by varying the viscosity contrast.

The importance of thermal fluctuations depends on the
the rotational Peclet number $Pe=\dot\gamma/D_r$.
The rotational diffusion coefficient $D_r$ is given by $D_r=k_B T/\zeta$
and employing the rotational friction coefficient $\zeta$ of a circle,
the Peclet number
can be written as $Pe=4 \pi \dot\gamma^* \kappa/(k_B T R_0)$.
In the following the Peclet number will
be changed by considering the values $\kappa/(k_B T R_0)=6.58, 65.8, 164.5$,
corresponding to $Pe=82, 821, 2041$, respectively, keeping fixed the
value of $\dot\gamma^*$.
The present study focuses on the dynamics and rheology
of a sheared vesicle
at finite values of $Pe$. Indeed, in previous studies of 
Refs.~\cite{ghig10,rahi10,zhao13,thie13,kaou14} it
was assumed $Pe=\infty$, thus neglecting the role of thermal
fluctuations.

%%%%%%%%%%%%%%%%%%%%%%%%%%%%%%%%%%%%%%%%%%
\section{Results}

We consider very dilute 
suspensions with a 
single vesicle 
for two values of the reduced area $S^*=0.80, 0.95$ corresponding effectively
to volume fractions $\phi=0.023, 0.028$, respectively.

\begin{figure}[H]
\includegraphics*[width=.47\textwidth]{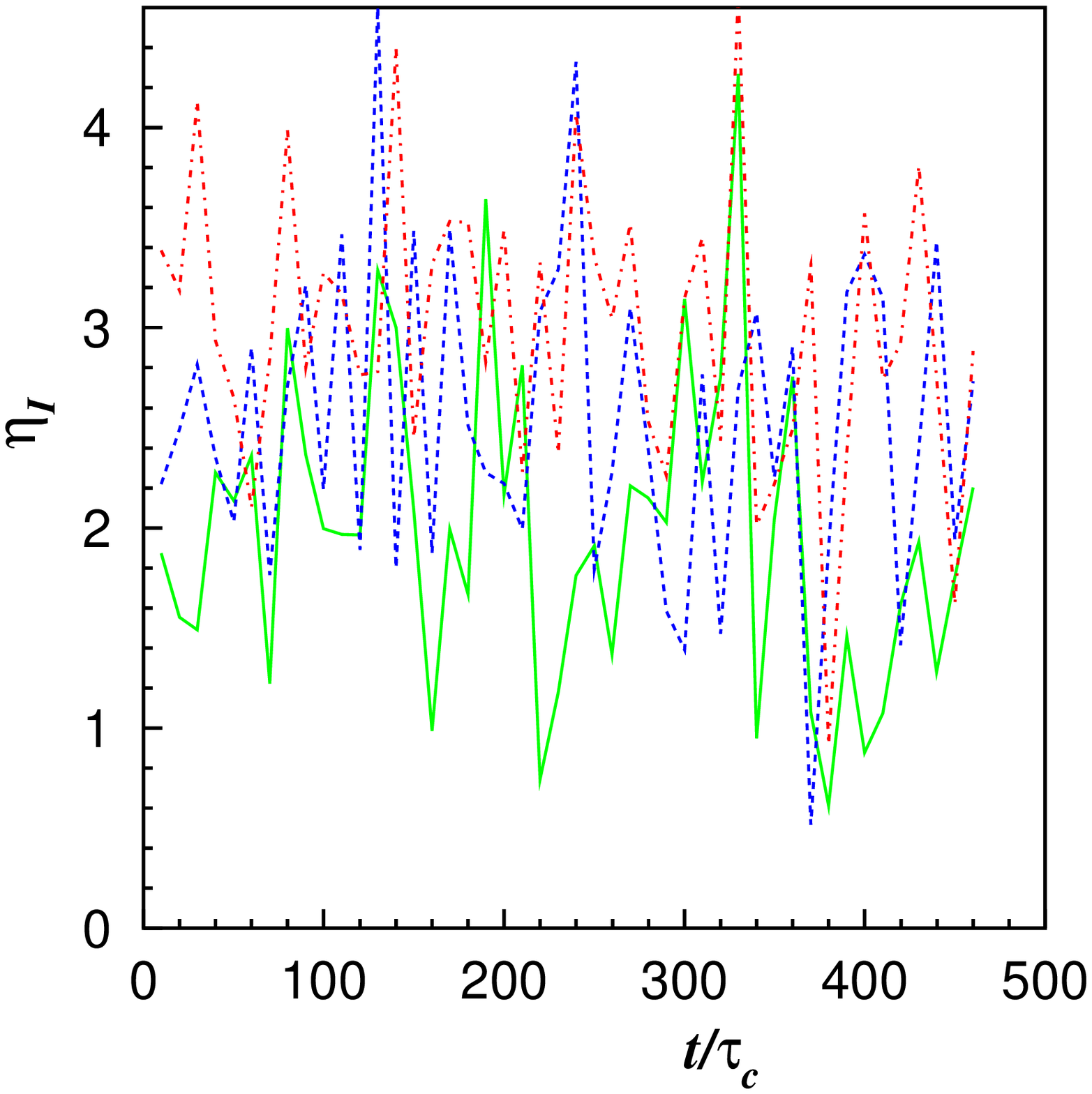}
\includegraphics*[width=.47\textwidth]{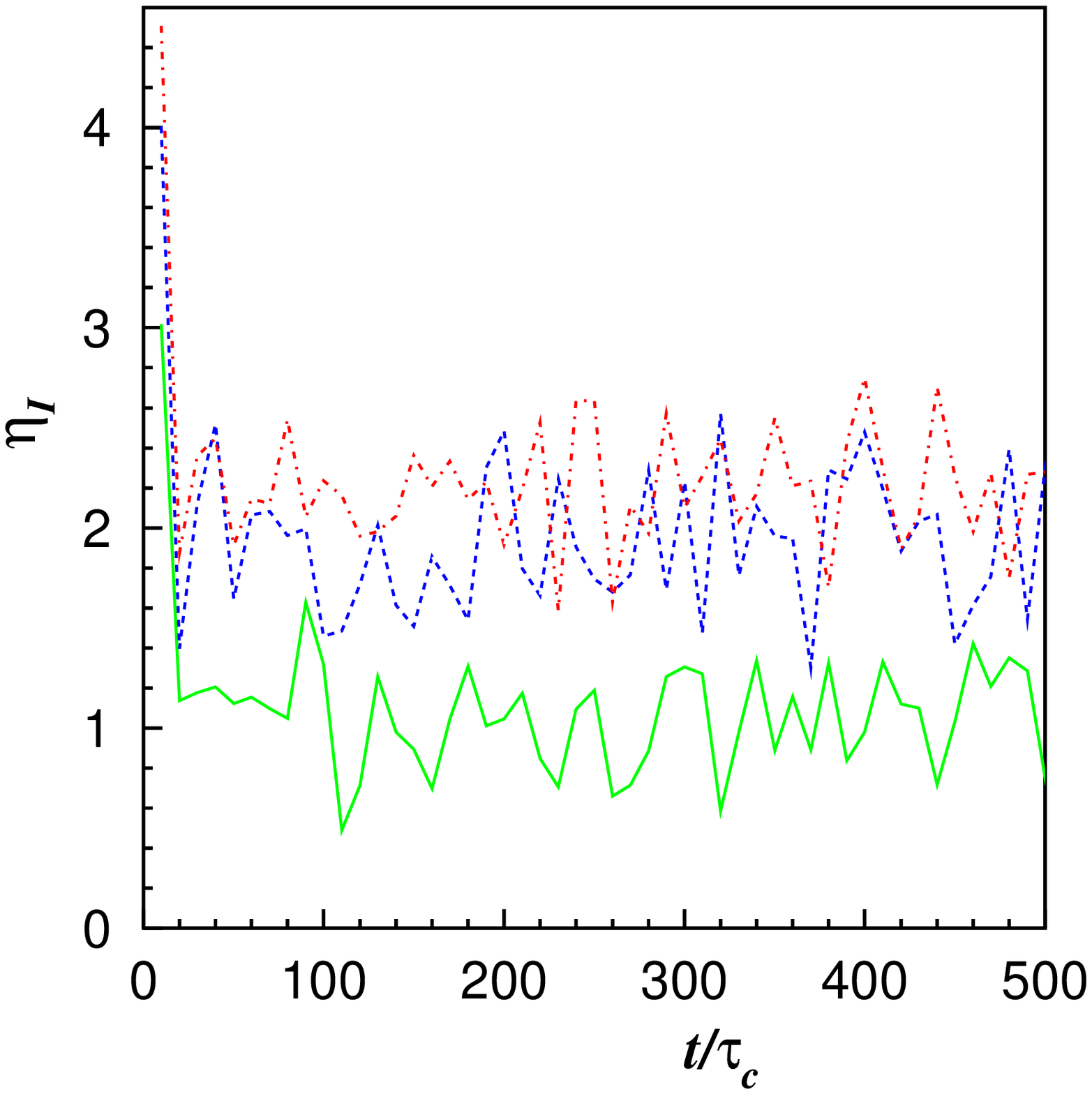}\\*
\includegraphics*[width=.47\textwidth]{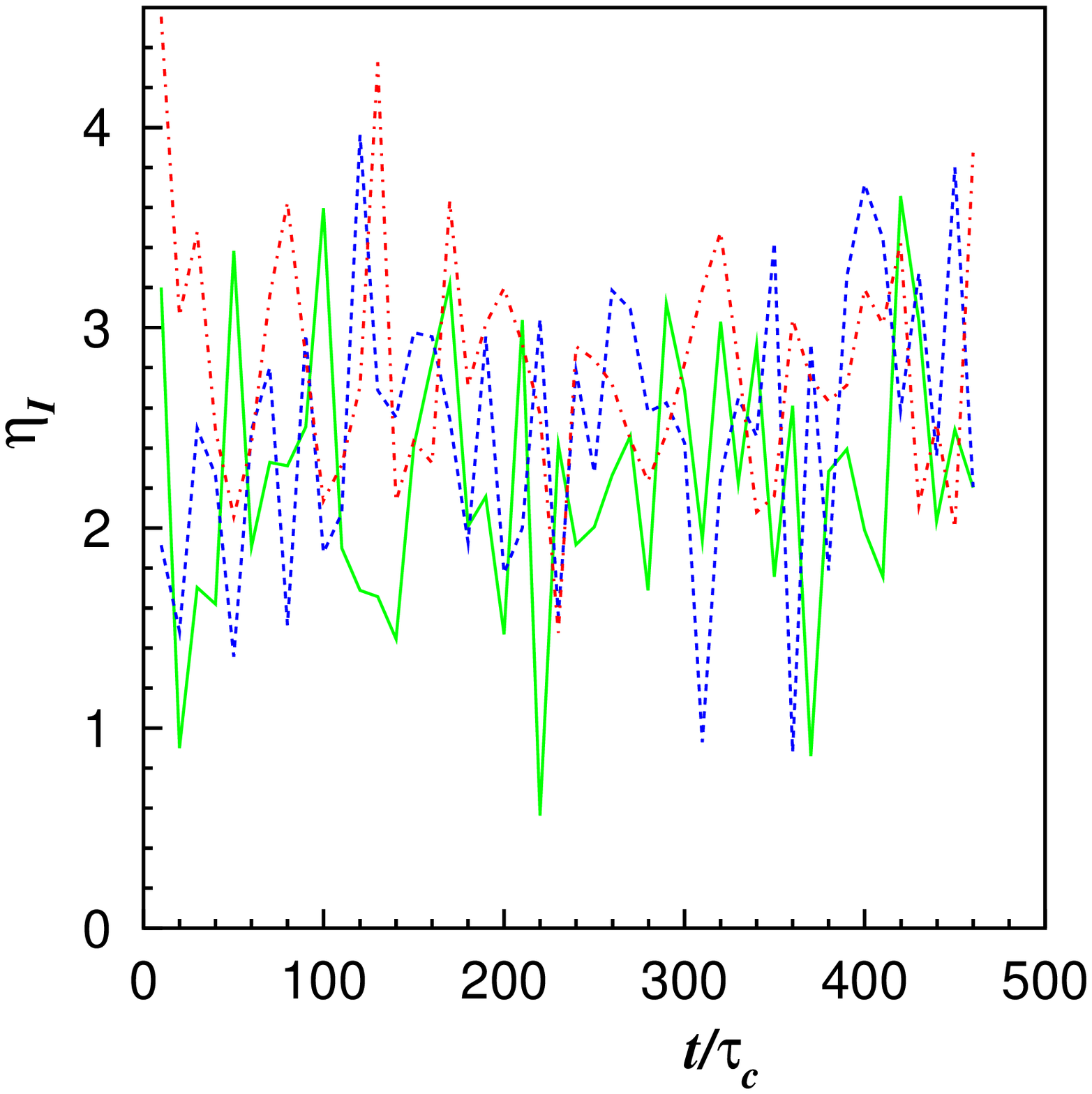}
\includegraphics*[width=.47\textwidth]{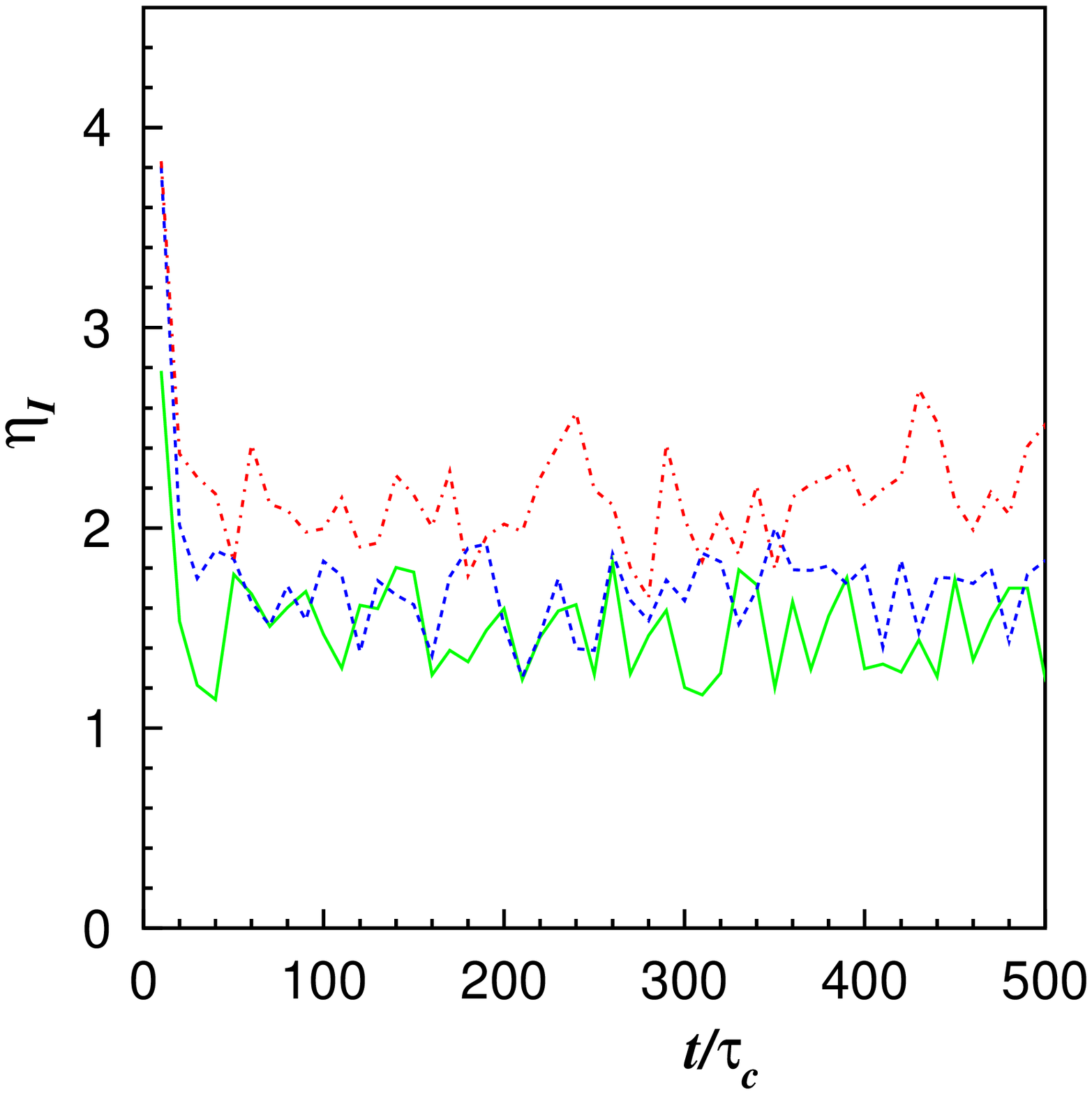}
\caption{Time behavior of the intrinsic viscosity $\eta_I$ (values are averaged
over time intervals of duration $\sim 10 \tau_c$ to smooth out noise)  
at $S^{*}=0.80$ (upper row) and 
$S^{*}=0.95$ (lower row)
for $\kappa/(k_B T R_0)=6.58$ (left), $65.8$ (right)
with $\lambda=$ 1 (full green line), 
7 (dashed blue line), 15 (dot-dashed red line).\label{fig1}}
\end{figure}

In Figure~\ref{fig1}, the instantaneous intrinsic viscosity 
$\eta_I$
is shown as a function of time for different values
of viscosity contrast $\lambda$, bending rigidity $\kappa$, 
and reduced area $S^*$. The viscosity $\eta$ is computed
as $\eta=\sigma_{xy}/\dot\gamma$ where $\sigma_{xy}$ is the $xy$
component of the stress tensor at walls \cite{mewi12}.
In the MPC model the stress $\sigma_{xy}$ has a contribution in the streaming
step, $\sigma_{xy}^s$, proportional to the flux of the $x$-momentum crossing
the walls, and a second contribution in the collision step, $\sigma_{xy}^c$,
due to the multi-particle collision with virtual wall particles
(see Appendix A).
 In two-dimensional simulations the streaming contribution is \cite{tao08}
\begin{equation}
  \sigma_{xy}^s=\frac{m}{L_x \Delta t_s} \sum_{i=1}^{N_s}
        [v'_{x,i}(t_b)-v_{x,i}(t_b)] ,
\end{equation}
where $t_b$ ($t \leq t_b \leq t+\Delta t_s$) is the time when particle $i$
bounces back from the wall, $v'_{x,i}(t_b)$ and $v_{x,i}(t_b)$ are the
velocities just after and before the collision with the wall, respectively,
and $N_s$
is the number of particles hitting one of the walls
in the time interval $[t,t+\Delta t_s]$.
The collision contribution is \cite{tao08}
\begin{equation}
\sigma_{xy}^c=\frac{m}{L_x \Delta t_s} \sum_{i=1}^{N_c}
        [v'_{x,i}(t+\Delta t_s)-v_{x,i}(t+\Delta t_s)] ,
\end{equation}
where $N_c$ is the number of particles having multi-particle collision with
virtual wall particles, while $v'_{x,i}(t+\Delta t_s)$
and $v_{x,i}(t+\Delta t_s)$ are the velocities of particle $i$ after and before
the collision step, respectively.

After a 
transient period, when the vesicle moves from the initial position towards 
the center of the
channel attaining its steady state, $\eta_I$ fluctuates around
average values up to the longest simulated times, which are more than two
orders of magnitude
larger than the vesicle relaxation time $\tau_c$. 

\begin{figure}[H]
\includegraphics*[width=.47\textwidth]{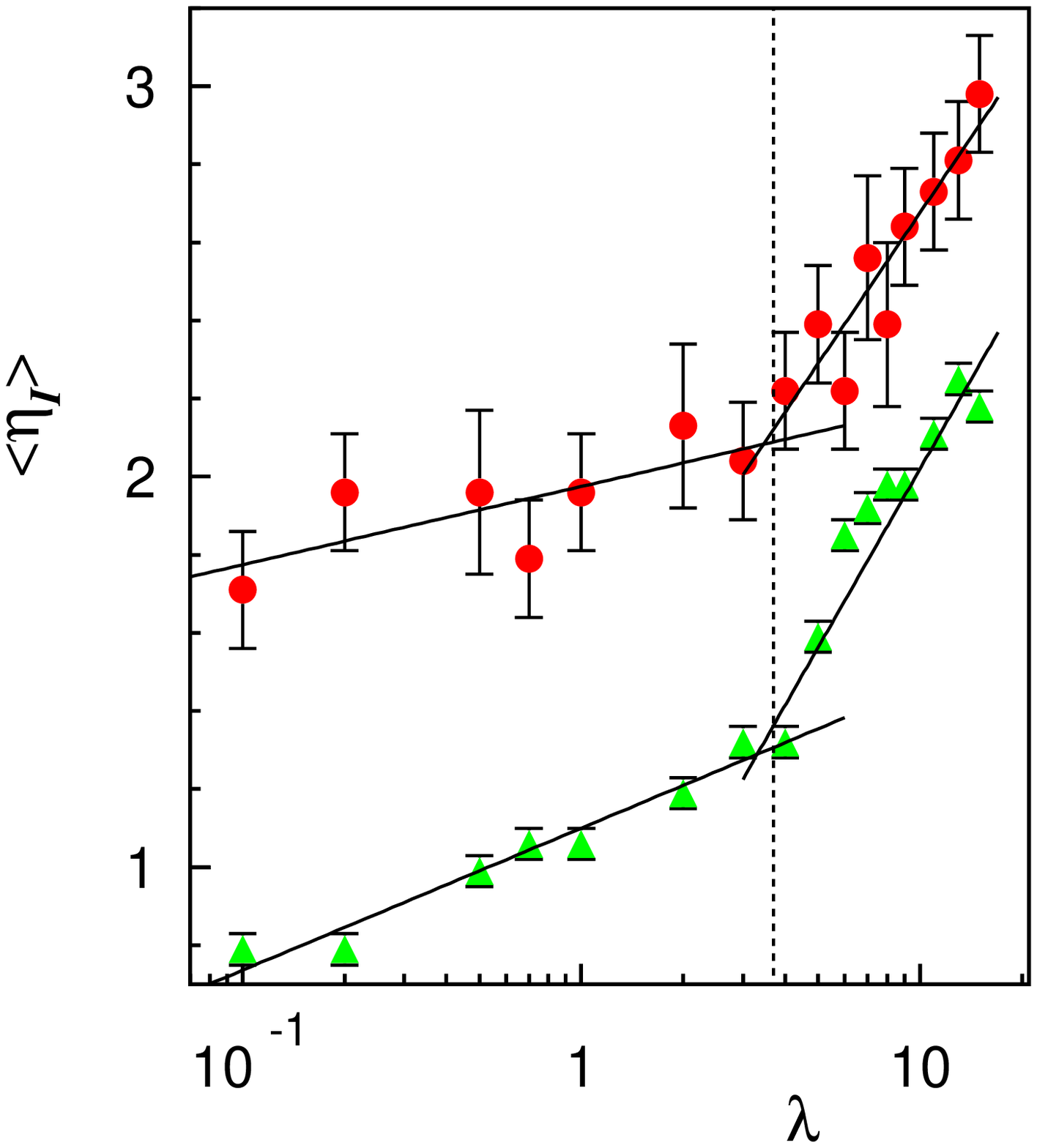}
\includegraphics*[width=.47\textwidth]{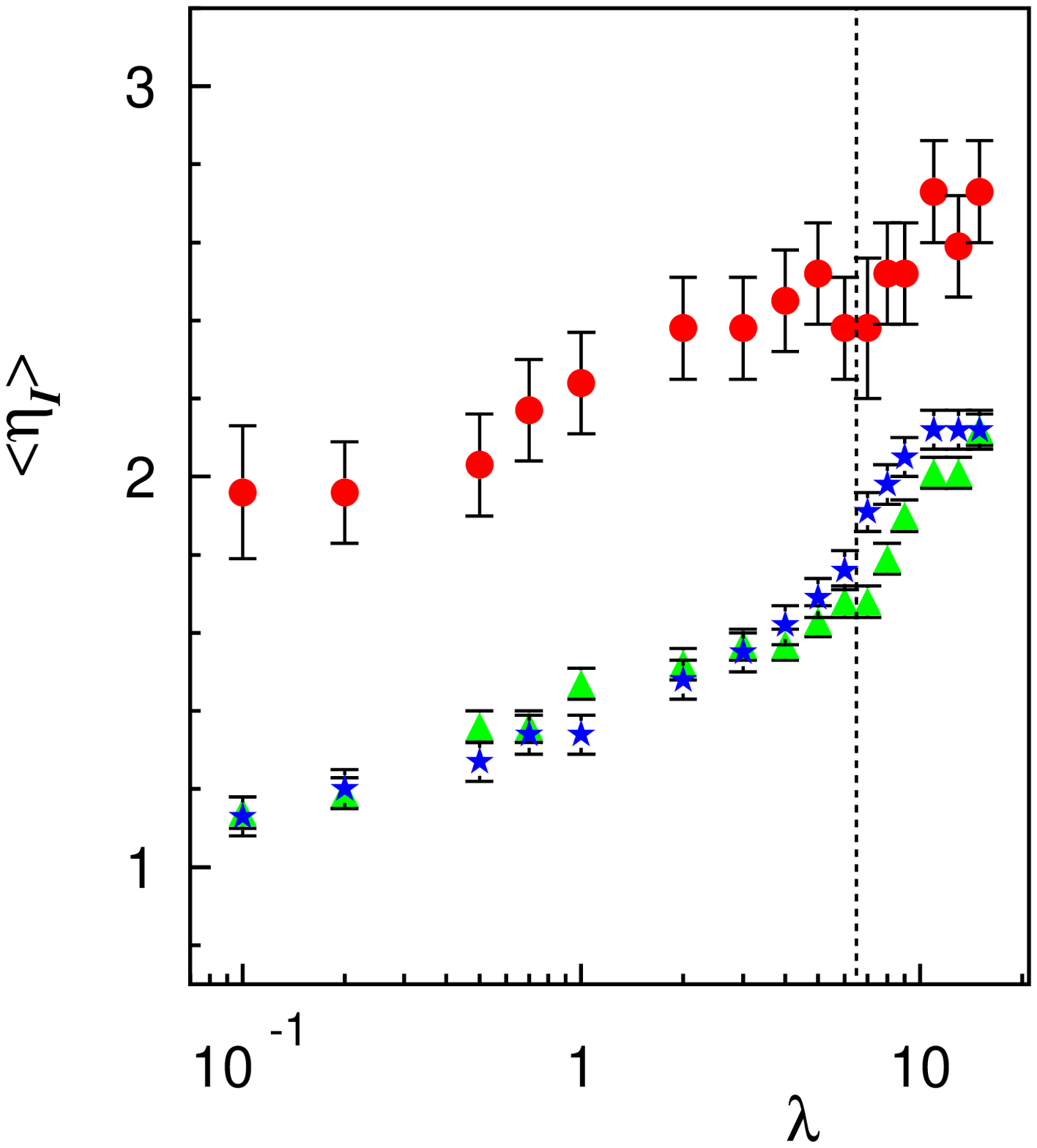}
\caption{Average values of the intrinsic viscosity $\langle \eta_I \rangle$ 
as a function of $\lambda$
at $S^{*}=0.80$ (left) and 
$S^{*}=0.95$ (right)
for $\kappa/(k_B T R_0)=6.58$ ($\bullet$), $65.8$ ($\blacktriangle$), $164.5$
($\star$). Full lines are guides to the eye.
The tank-treading to tumbling transition occurs at $\lambda_c \simeq 3.7$
for $S^{*}=0.80$ and at $\lambda_c \simeq 6.5$
for $S^{*}=0.95$ in the Keller-Skalak theory \cite{kell82} and is marked by the dashed
vertical lines. Error bars are given by the root-mean-square fluctuation
values  of the intrinsic viscosity.
\label{fig2}}
\end{figure}

The values $\langle \eta_I \rangle$ of the intrinsic viscosity, 
time-averaged in the
steady state, are reported in Figure~\ref{fig2} as 
a function of $\lambda$.
It appears that $\langle \eta_I \rangle$ is an increasing function of $\lambda$ for
the used values of the reduced area, bending energy, and 
temperature, in 
agreement with our previous results \cite{lamu13,lamu15}.
In the Keller-Skalak theory \cite{kell82}, where thermal fluctuations are ignored,
the sharp TT-to-TU transition occurs at $\lambda_c \simeq 3.7$
for $S^{*}=0.80$ and at $\lambda_c \simeq 6.5$
for $S^{*}=0.95$. However, finite temperature broadens the
TT-to-TU transition \cite{mess09}.
In the TU regime at higher values of $\lambda$, 
the growth of $\langle \eta_I \rangle$ is steeper.
A decrease of the intrinsic viscosity 
in the TT regime followed by its growth in the TU regime, as
theoretically predicted in Refs.~\cite{dank07a,dank07b} 
and observed in simulations 
without thermal fluctuations \cite{ghig10,rahi10,zhao13,thie13,kaou14}, 
is not found in our model. The effect of increasing the bending energy
is to reduce the value of the intrinsic viscosity without changing
the monotonic dependence on the viscosity contrast. 
This effect seems to be triggered by the Peclet number
as it will be later discussed.

\begin{figure}[H]
\includegraphics*[width=.47\textwidth]{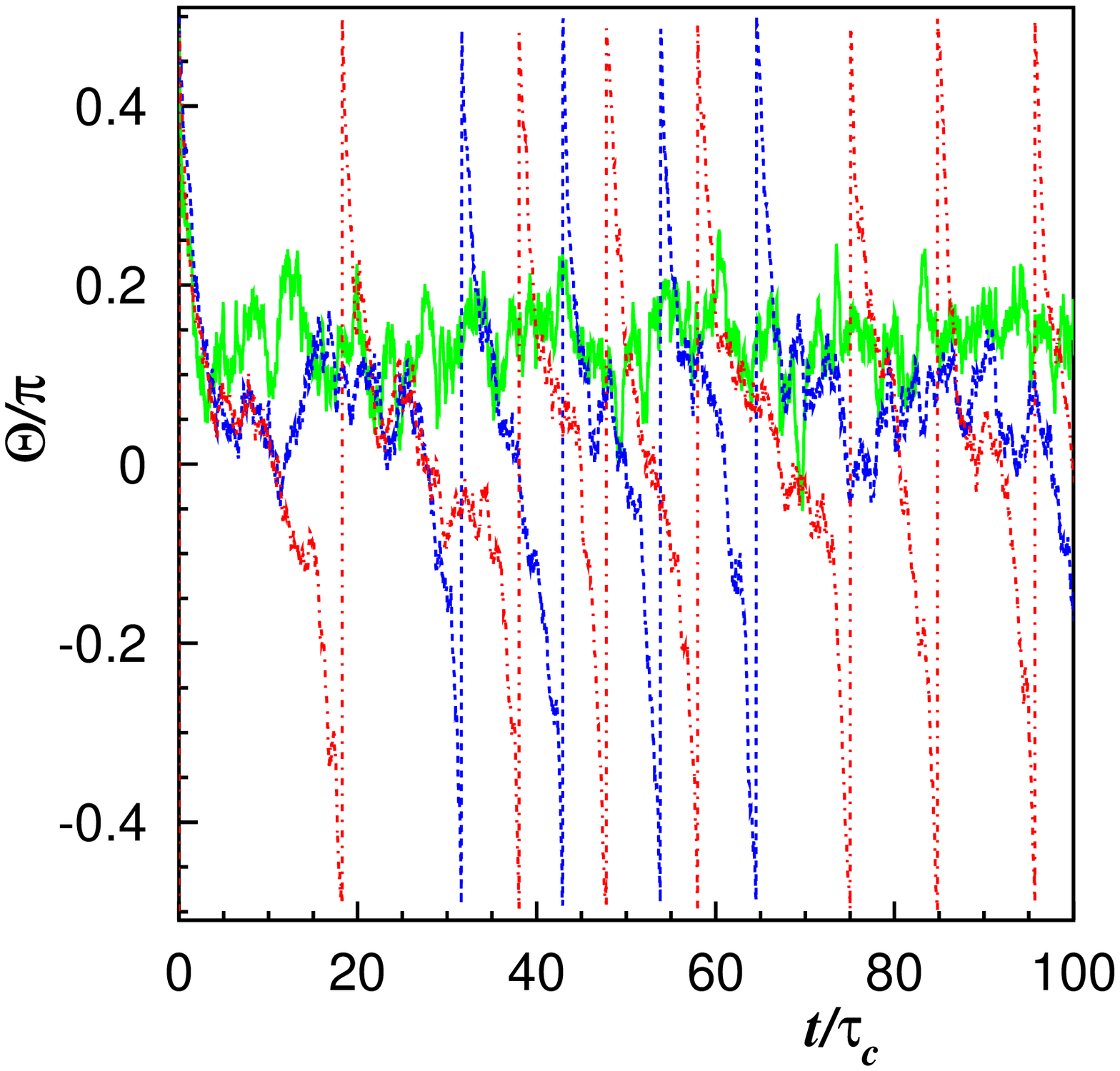}
\includegraphics*[width=.47\textwidth]{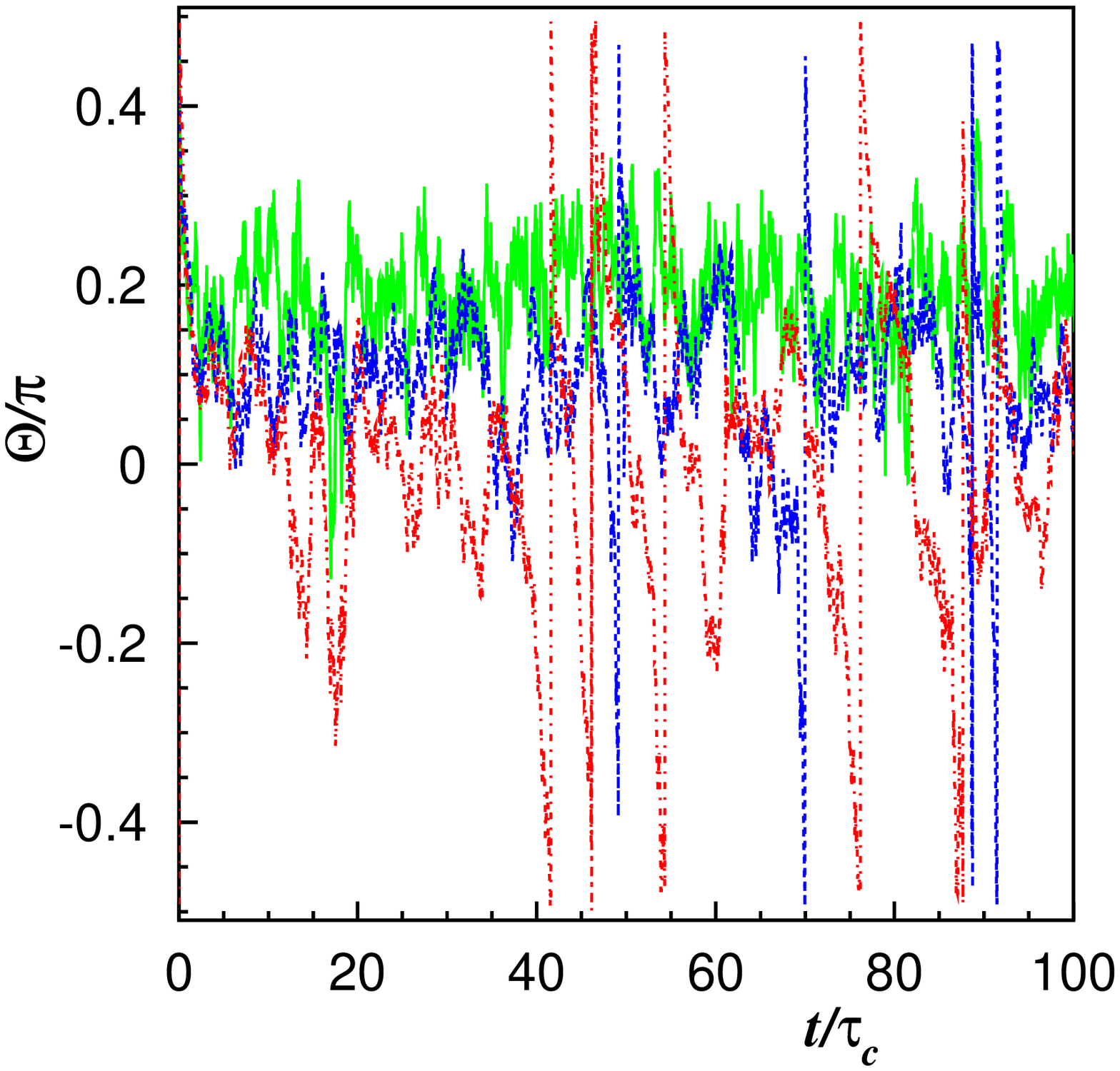}
\caption{Time behavior of the inclination angle $\Theta$ 
at $S^{*}=0.80$ (left) and 
$S^{*}=0.95$ (right) 
for $\kappa/(k_B T R_0)=6.58$ with $\lambda=$ 1 (full green line), 
7 (dashed blue line), 15 (dot-dashed red line).\label{fig3}}
\end{figure}

\begin{figure}[H]
\includegraphics*[width=.47\textwidth]{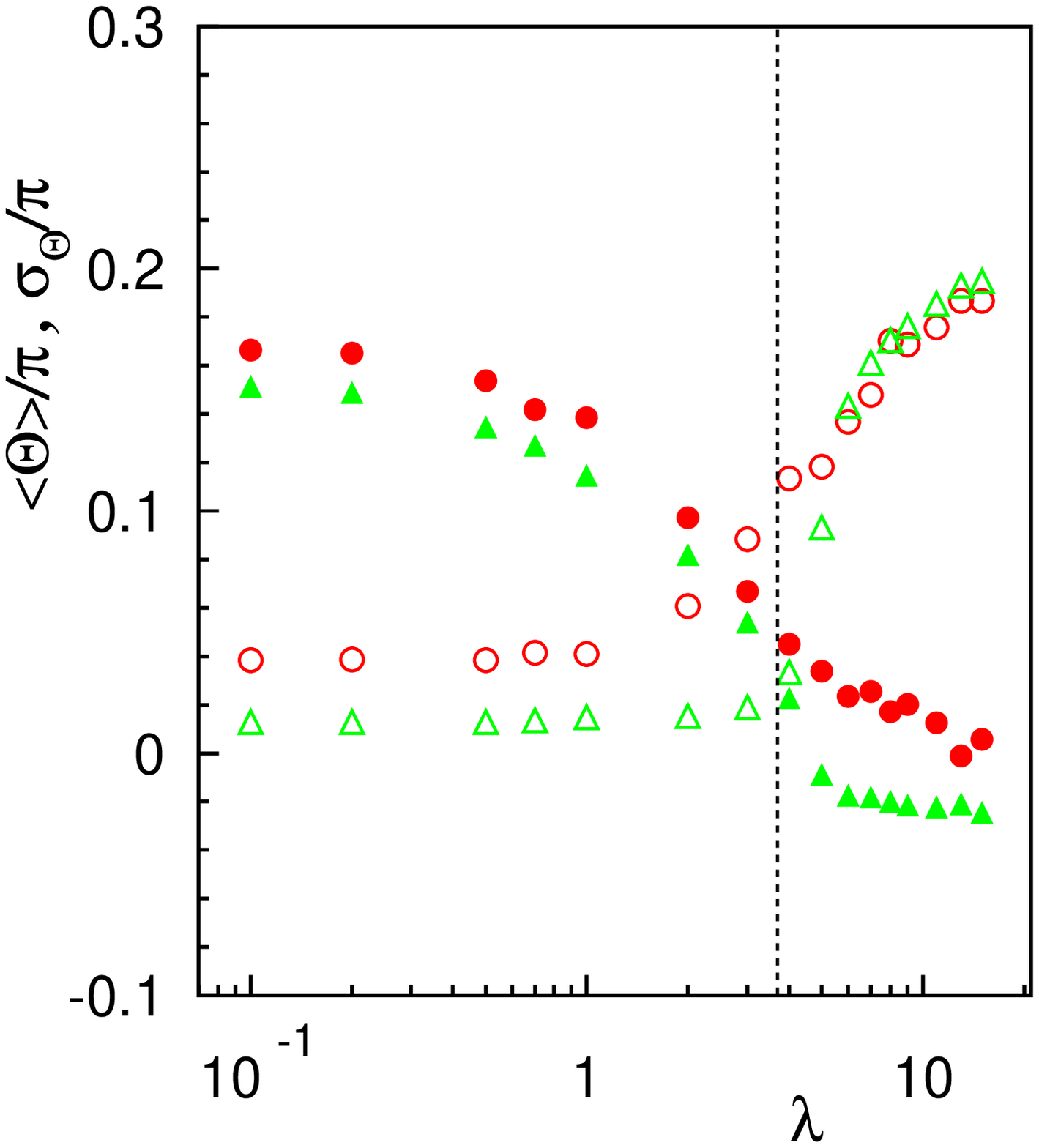}
\includegraphics*[width=.47\textwidth]{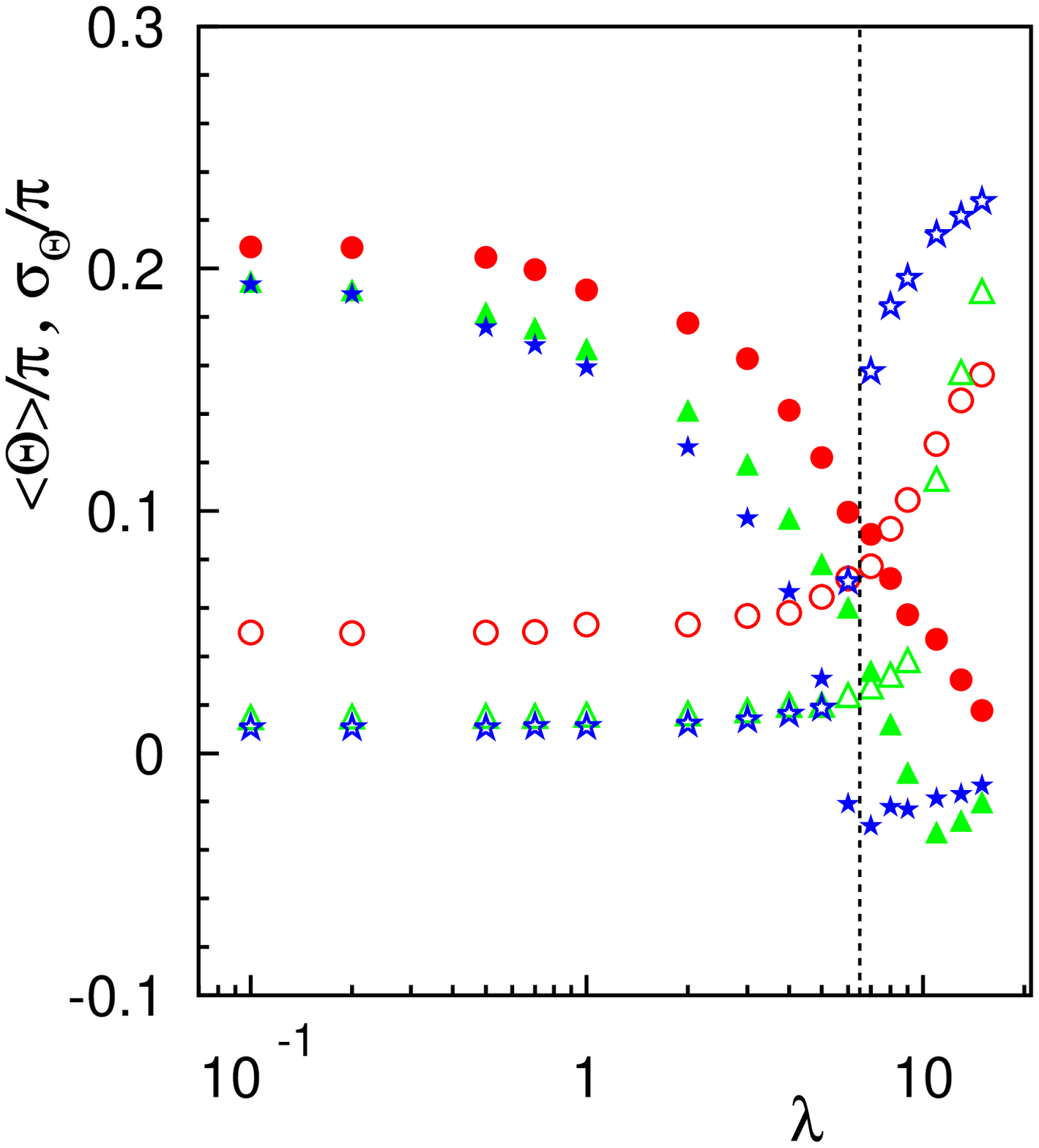}
\caption{Average values of the inclination angle $\langle \Theta \rangle$ 
(filled symbols)
and its rms fluctuation values $\sigma_{\Theta}$ 
(empty symbols) as a function of $\lambda$
at $S^{*}=0.80$ (left) and 
$S^{*}=0.95$ (right)
for $\kappa/(k_B T R_0)=6.58$ ($\bullet$), $65.8$ ($\blacktriangle$), $164.5$
($\star$). The tank-treading to tumbling transition
in the Keller-Skalak theory \cite{kell82} is marked by the dashed
vertical lines.\label{fig4}}
\end{figure}

In order to clarify the observed behavior of $\langle \eta_I \rangle$, 
the vesicle dynamics was investigated in more detail by monitoring
the temporal evolution of several
quantities. The inclination angle $\Theta$, describing 
the angle between the $x$ direction
and the long main axis of the vesicle, can be used to discriminate
between the TT and the TU states. In the former case, $\Theta$ 
reaches a steady value, while in the latter case, $\Theta$ varies periodically 
in time.
In Figure~\ref{fig3}, the inclination angle is shown as a function of time. 
For low values of $\lambda$ the vesicle performs tank-treading motion
and the inclination angle fluctuates around a steady value.
In contrast, without thermal fluctuations 
\cite{thie13,kaou14}
the inclination angle is constant in the TT regime after the initial transient. 
When increasing the viscosity contrast, some tumbling events appear, which
become predominant for the highest value of $\lambda$.
The time-averaged values $\langle \Theta \rangle$ are depicted
in Figure~\ref{fig4},  together with the root-mean-square (rms) fluctuation values 
$\sigma_{\Theta}=\sqrt{\langle (\Delta \Theta)^2 \rangle}$.
The transition from the TT to the TU regime, which is characterized by
going from values
$\langle \Theta \rangle > 0$ to $\langle \Theta \rangle \simeq 0$, 
is broader for the smallest values of the bending rigidity, 
and gets sharper when increasing the ratio $\kappa/(k_B T R_0)$.
The fluctuations $\sigma_{\Theta}$ reduce with $Pe$ in the TT regime, as 
theoretically predicted \cite{fink08}, 
and show an opposite trend with increasing
viscosity contrast.

From the gyration tensor of the vesicle, the two eigenvalues $\Lambda_M$
and $\Lambda_m$ with $\Lambda_M > \Lambda_m$ are extracted and the 
asphericity $A=[(\Lambda_M - \Lambda_m)/(\Lambda_M + \Lambda_m)]^2$
is computed.
\begin{figure}[H]
\includegraphics*[width=.47\textwidth]{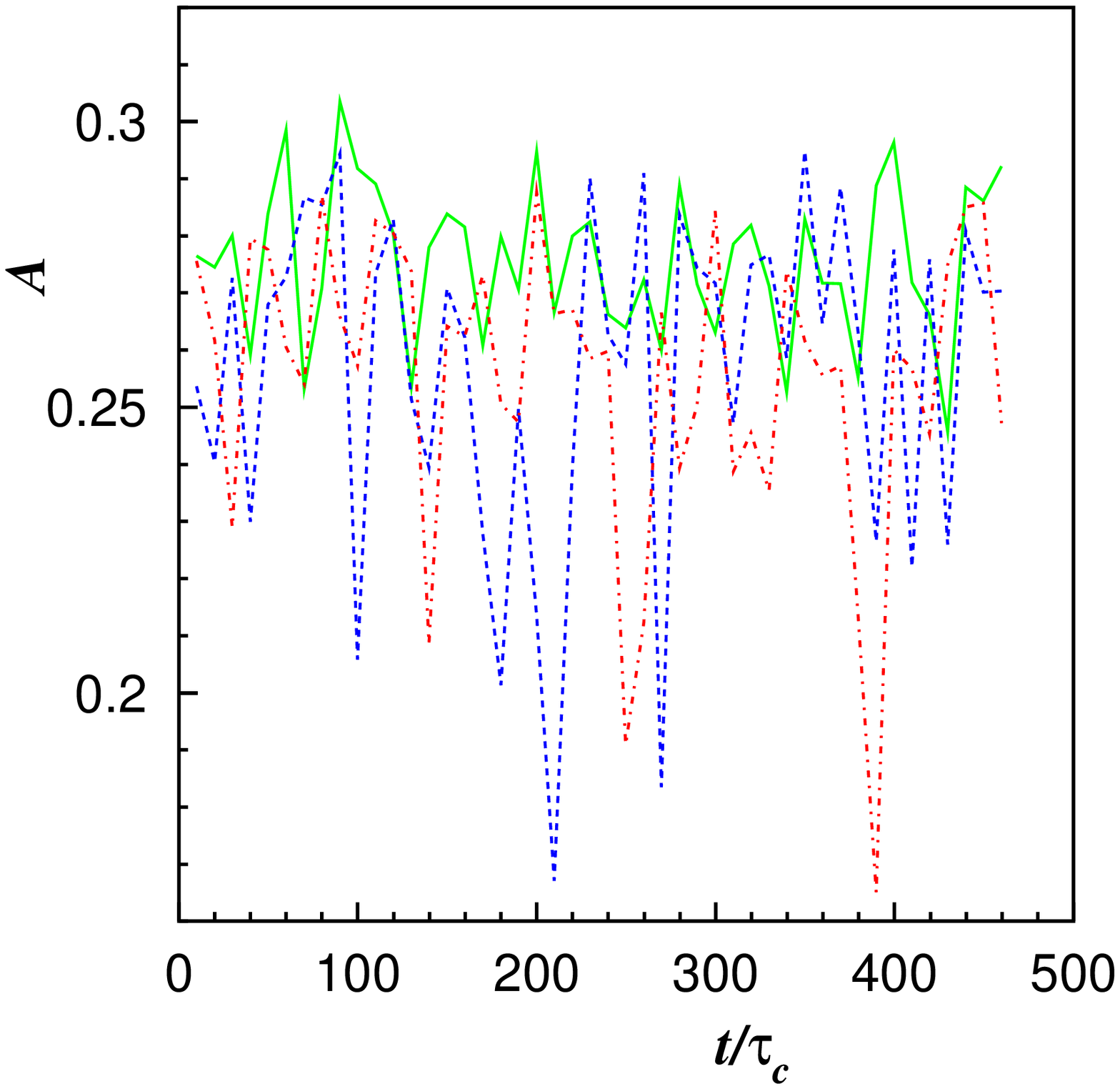}
\includegraphics*[width=.47\textwidth]{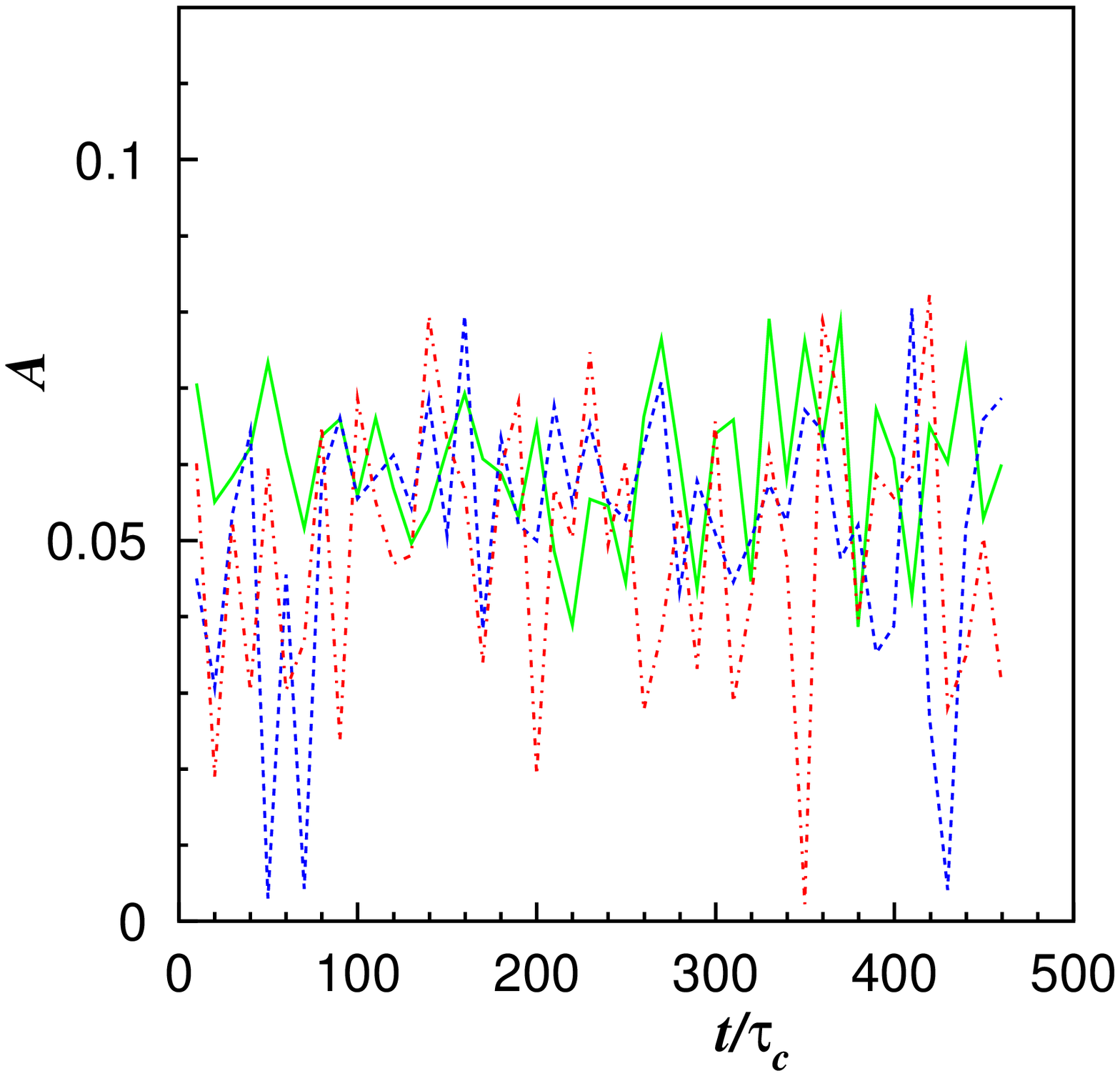}
\caption{Time behavior of the asphericity $A$ (values are sampled every 
$10 \tau_c$) at $S^{*}=0.80$ (left) and 
$S^{*}=0.95$ (right)
for $\kappa/(k_B T R_0)=6.58$ with $\lambda=$ 1 (full green line), 
7 (dashed blue line), 15 (dot-dashed red line).\label{fig5}} 
\end{figure}
The values of $A$ as a function of time are shown in Figure~\ref{fig5}
and the time-averages $\langle A \rangle$ as a function of the viscosity contrast
in Figure~\ref{fig6}.
\begin{figure}[H]
\includegraphics*[width=.47\textwidth]{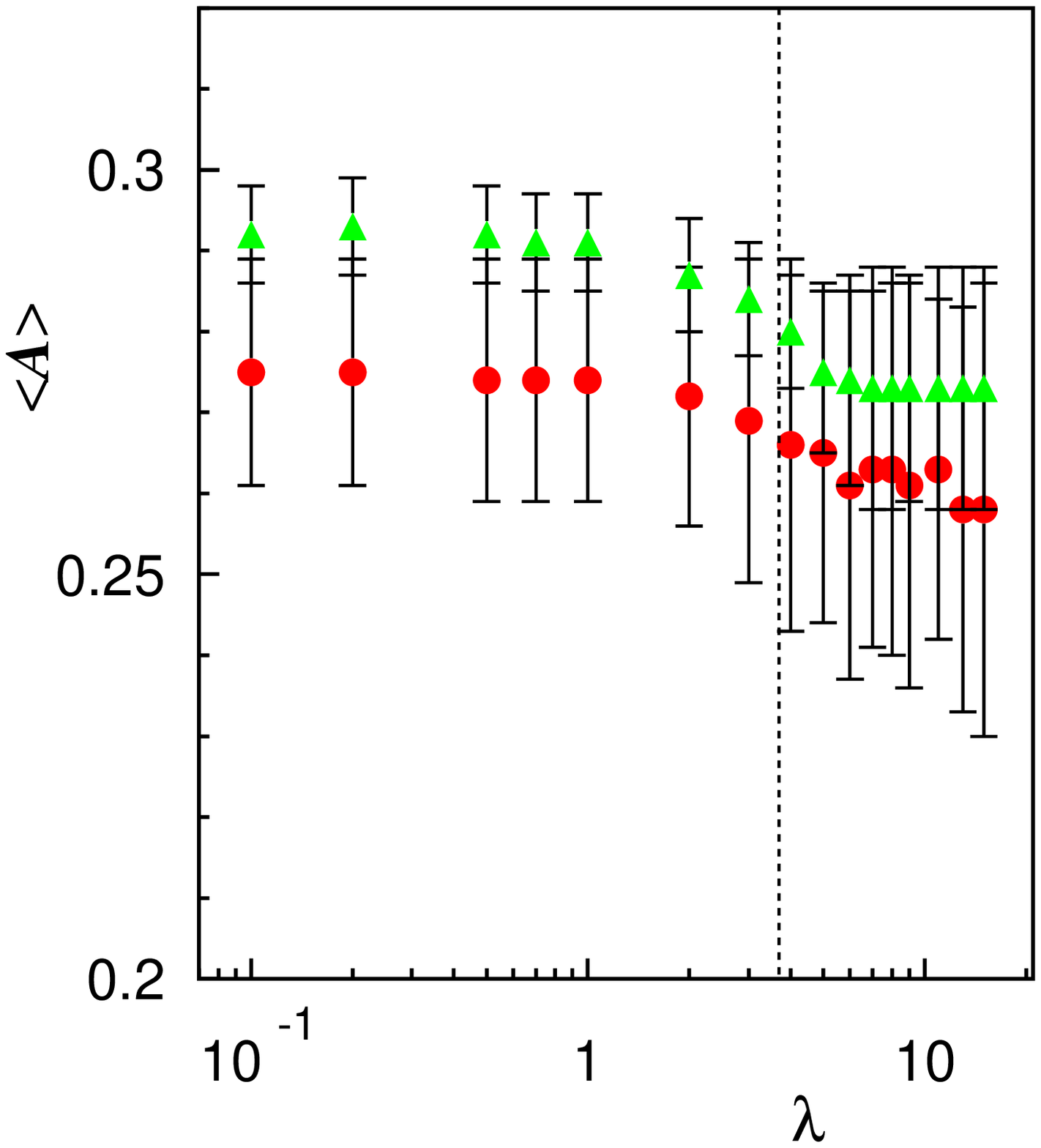}
\includegraphics*[width=.47\textwidth]{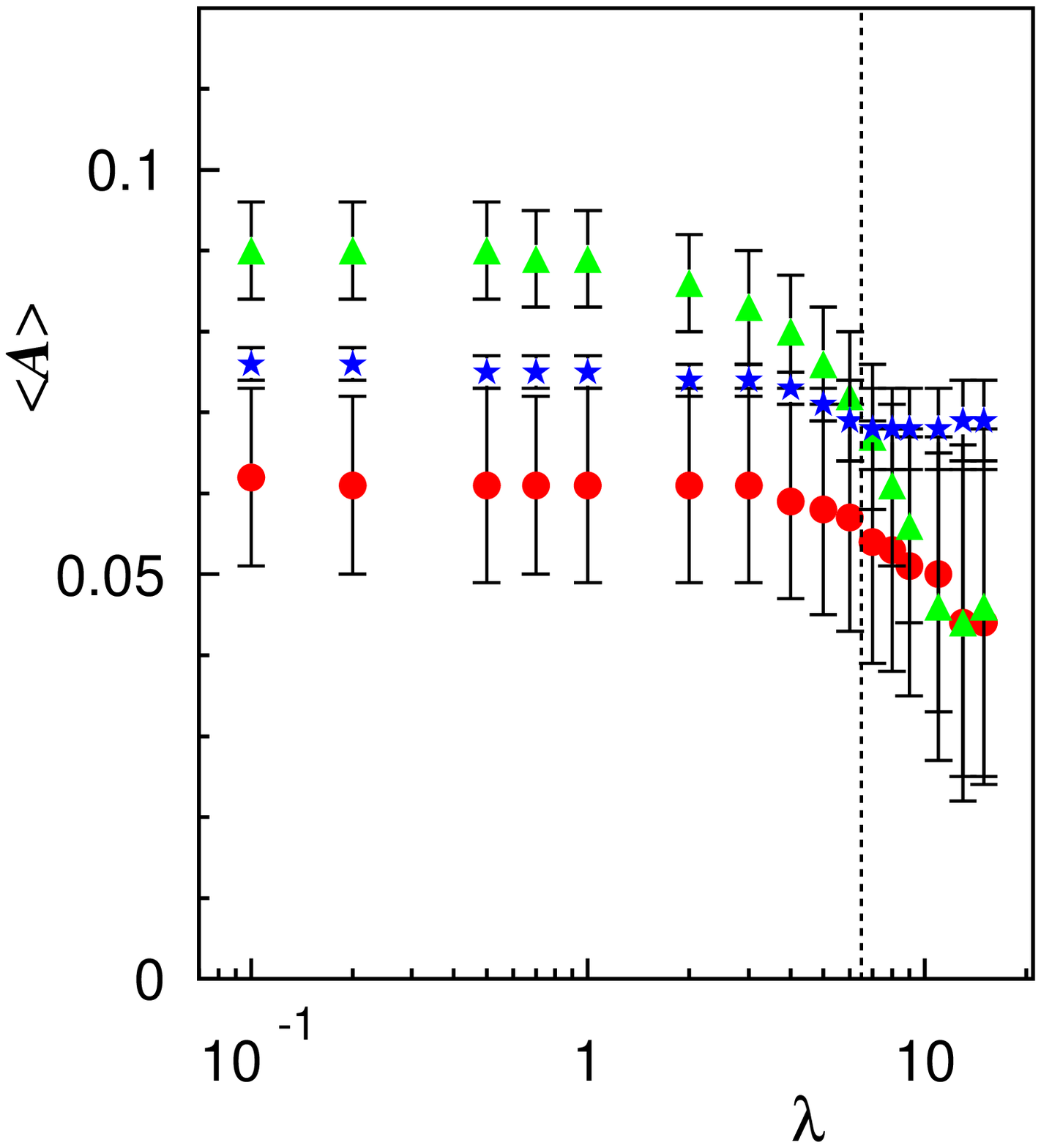}
\caption{Average values of the asphericity $\langle A \rangle$ 
as a function of $\lambda$
at $S^{*}=0.80$ (left) and 
$S^{*}=0.95$ (right)
for $\kappa/(k_B T R_0)=6.58$ ($\bullet$), $65.8$ ($\blacktriangle$), $164.5$
($\star$). The tank-treading to tumbling transition
in the Keller-Skalak theory \cite{kell82} is marked by the dashed
vertical lines. Error bars are given by the root-mean-square fluctuation
values  of the asphericity.
\label{fig6}}
\end{figure}
$\langle A \rangle$ is constant in the TT regime and 
decreases when approaching the TU regime, 
showing that the vesicle becomes more rounded when the inner fluid 
is more viscous. Also, $\langle A \rangle$ is smaller for the lower value
of bending rigidity 
and does not change significantly going from TT to TU
regime for the highest value of the bending rigidity. 
In the case of the quasi-circular vesicle a non-monotonic behavior
of $\langle A \rangle$ with the bending rigidity can be observed in
the TT regime.  
The average values $\langle \sqrt{\Lambda_M} \rangle$ and 
$\langle \sqrt{\Lambda_m} \rangle$, which
give an estimate of the vesicle semi-axes, are
plotted in Figure~\ref{fig7} as a function of the viscosity contrast to demonstrate
how the vesicle becomes more rounded when increasing $\lambda$
for fixed $\dot \gamma^*$.
\begin{figure}[H]
\includegraphics*[width=.5\textwidth]{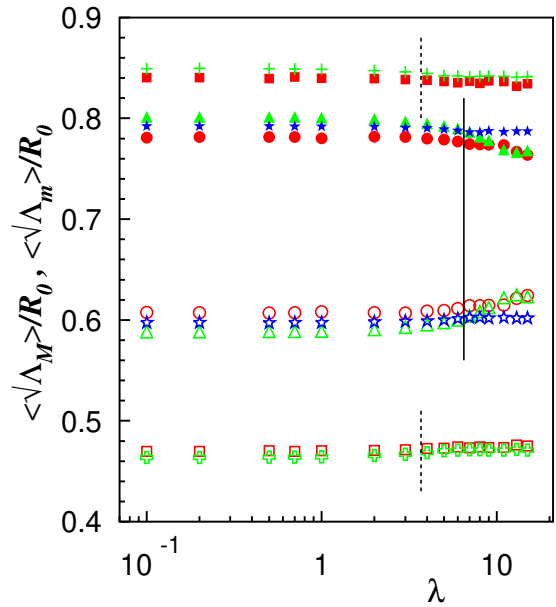}
\caption{Average values of the gyration tensor eigenvalues 
$\langle \sqrt{\Lambda_M} \rangle$ 
(filled symbols) and $\langle \sqrt{\Lambda_m} \rangle$ (empty symbols)
as a function of $\lambda$
at $S^{*}=0.80$ 
for $\kappa/(k_B T R_0)=6.58$ ($\Box$), $65.8$ ($+$), and 
at $S^{*}=0.95$
for $\kappa/(k_B T R_0)=6.58$ ($\bullet$), $65.8$ ($\blacktriangle$), $164.5$
($\star$). The tank-treading to tumbling transition
in the Keller-Skalak theory \cite{kell82} is marked by the 
dashed ($S^{*}=0.80$) and full ($S^{*}=0.95$)
vertical lines.\label{fig7}
}
\end{figure}
It can be seen that $\langle \sqrt{\Lambda_M} \rangle$ decreases and 
$\langle \sqrt{\Lambda_m} \rangle$
increases as functions of $\lambda$. The relative change of the average
eigenvalues, going from
the TT to the TU regime, is larger at $\kappa/(k_B T R_0)=65.8$ while
it is negligible for the highest value of the bending rigidity.
The rms fluctuation values $\sigma_M=\sqrt{\langle (\Delta \sqrt{\Lambda_M})^2 \rangle}$
and $\sigma_m=\sqrt{\langle (\Delta \sqrt{\Lambda_m})^2 \rangle}$ are reported
in Figure~\ref{fig8} as functions of $\lambda$. 
In all the cases the values of the rms fluctuations are constant in the TT
regime and increase when 
entering the TU regime. Moreover, $\sigma_M$ and $\sigma_m$
decrease when increasing the Peclet number.
\begin{figure}[H]
\includegraphics*[width=.47\textwidth]{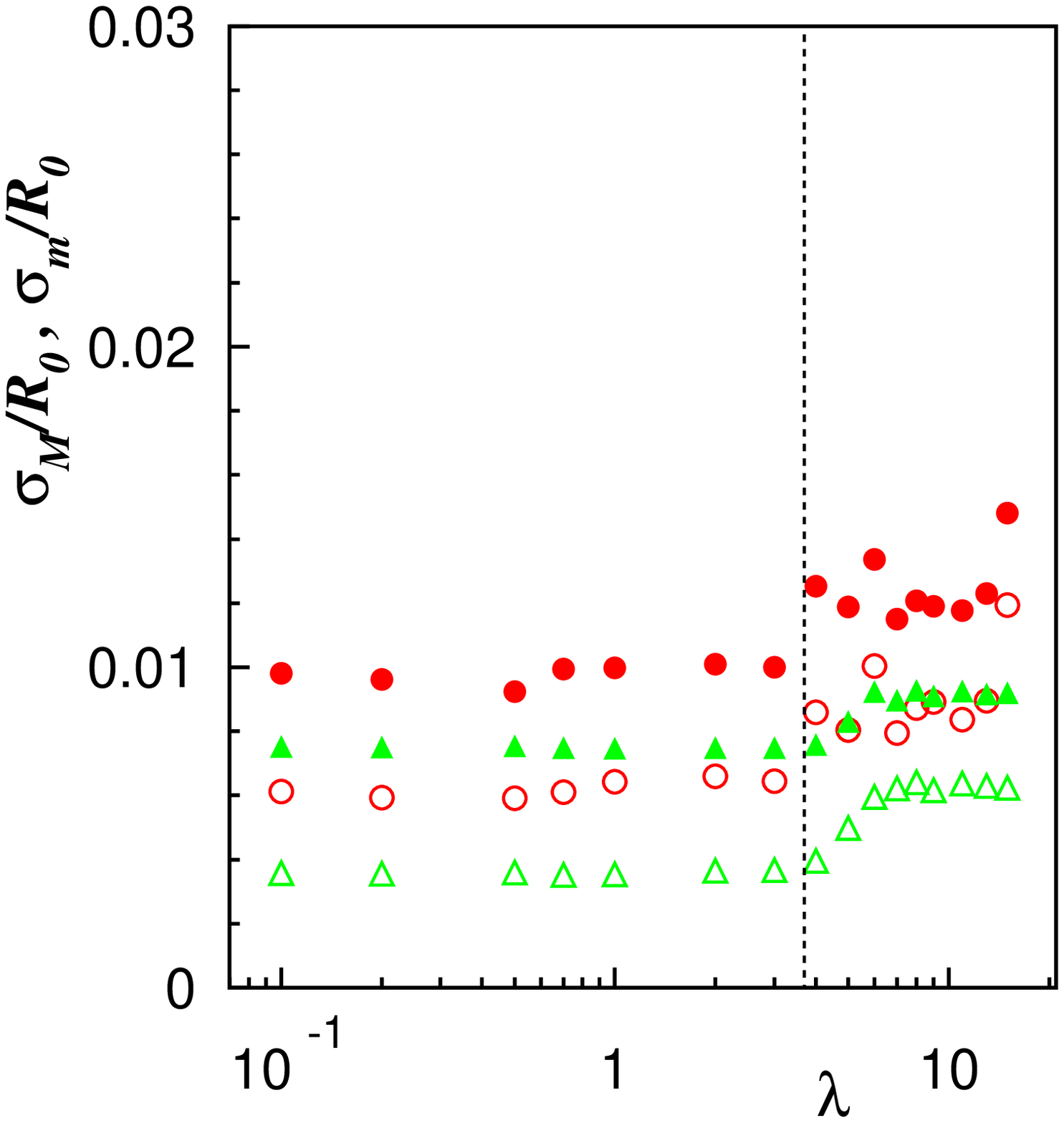}
\includegraphics*[width=.47\textwidth]{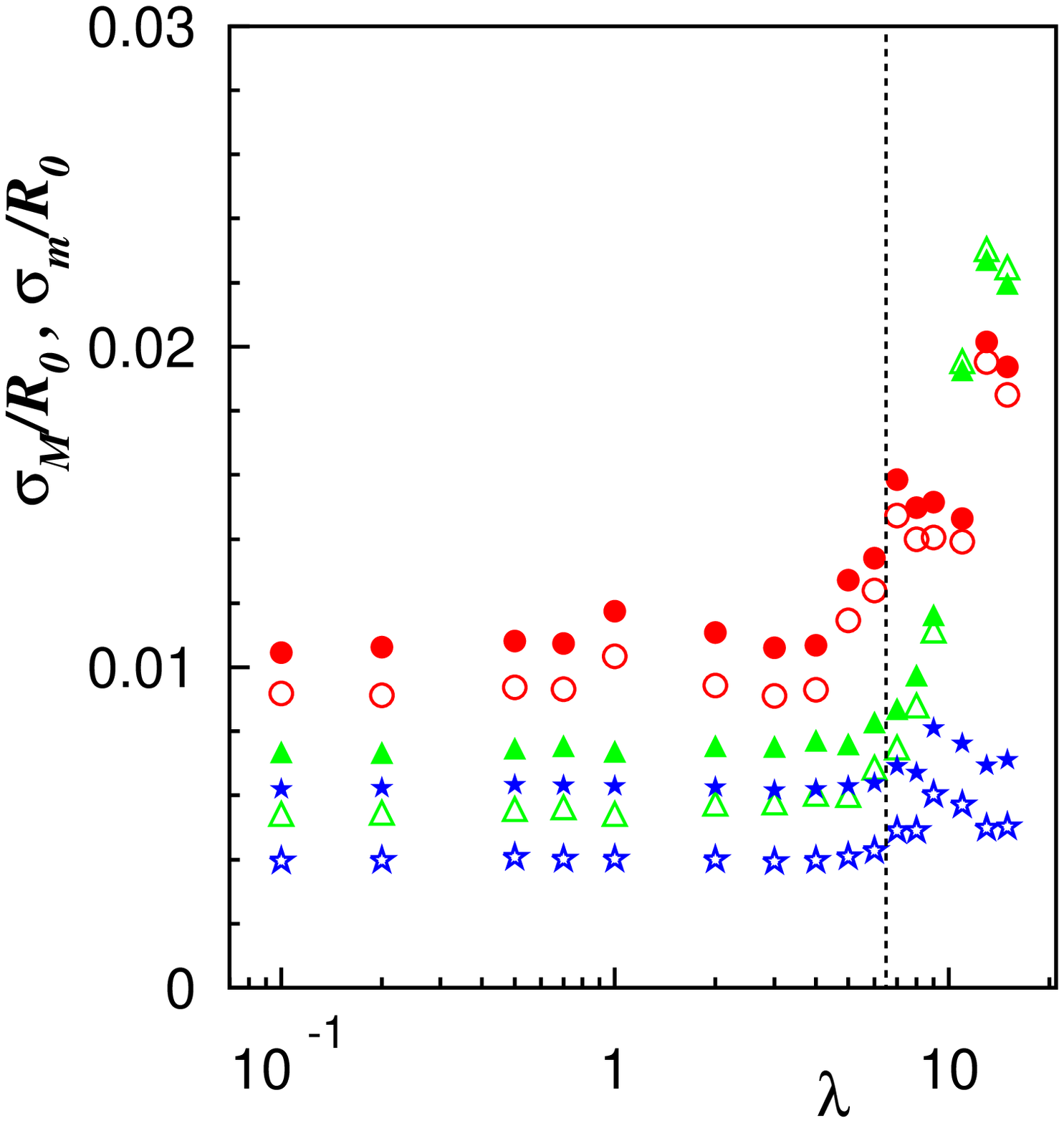}
\caption{Rms fluctuation values 
$\sigma_M$ 
(filled symbols) and $\sigma_m$ (empty symbols)
of the the gyration tensor eigenvalues of Fig.~7
as a function of $\lambda$
at $S^{*}=0.80$ (left) and 
$S^{*}=0.95$ (right)
for $\kappa/(k_B T R_0)=6.58$ ($\bullet$), $65.8$ ($\blacktriangle$), $164.5$
($\star$). The tank-treading to tumbling transition
in the Keller-Skalak theory \cite{kell82} is marked by the dashed
vertical lines.\label{fig8}}
\end{figure}

\begin{figure}[H]
\includegraphics*[width=.47\textwidth]{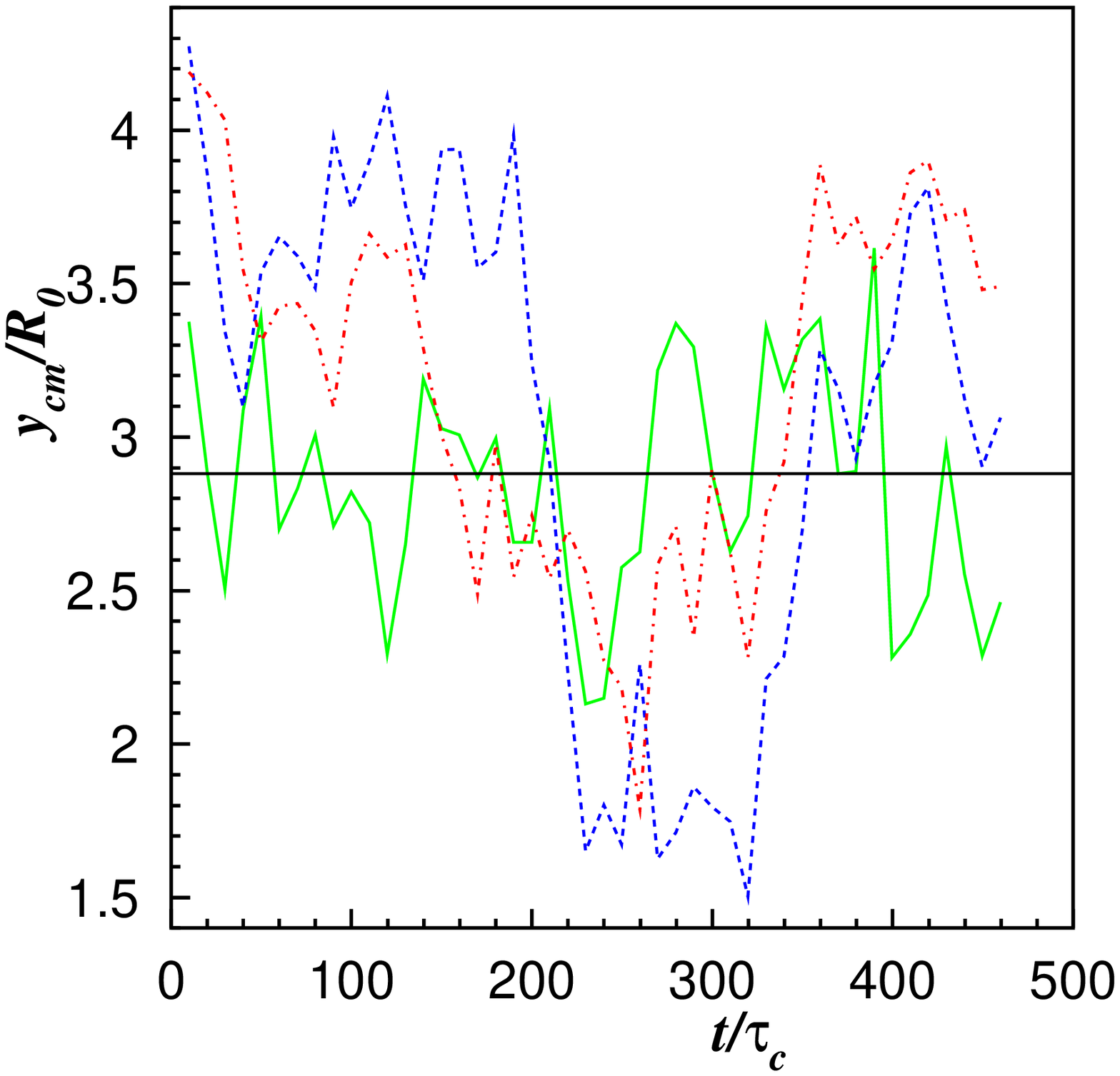}
\includegraphics*[width=.47\textwidth]{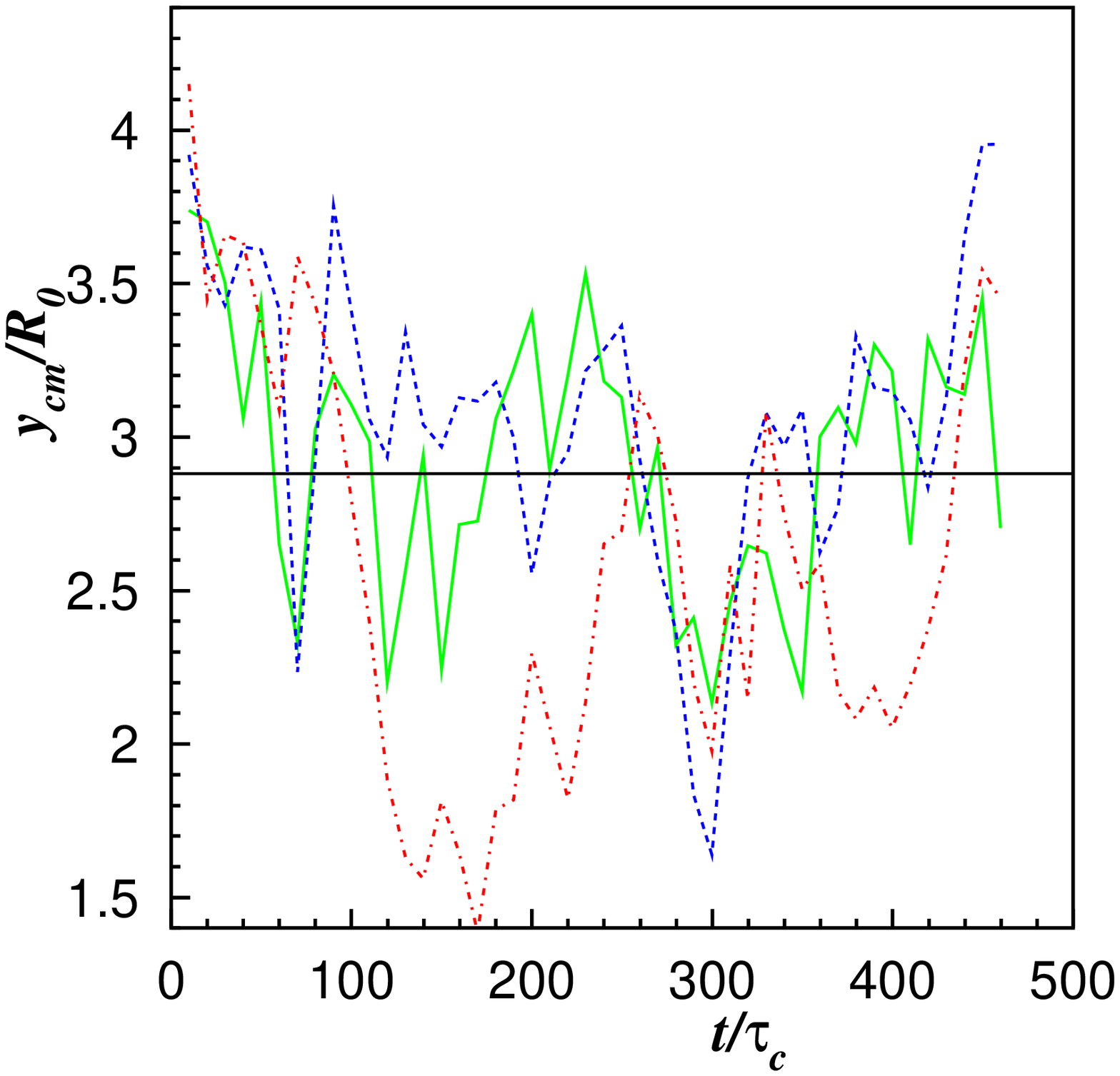}
\caption{Time behavior of the vertical position $y_{cm}$ of the vesicle center
of mass (values are sampled every 
$10 \tau_c$) at $S^{*}=0.80$ (left) and 
$S^{*}=0.95$ (right)
for $\kappa/(k_B T R_0)=6.58$ 
with $\lambda=$ 1 (full red line), 
7 (dashed blue line), 15 (dot-dashed red line). 
The horizontal full line denotes the center of the channel.\label{fig9}} 
\end{figure}

\begin{figure}[H]
\includegraphics*[width=.47\textwidth]{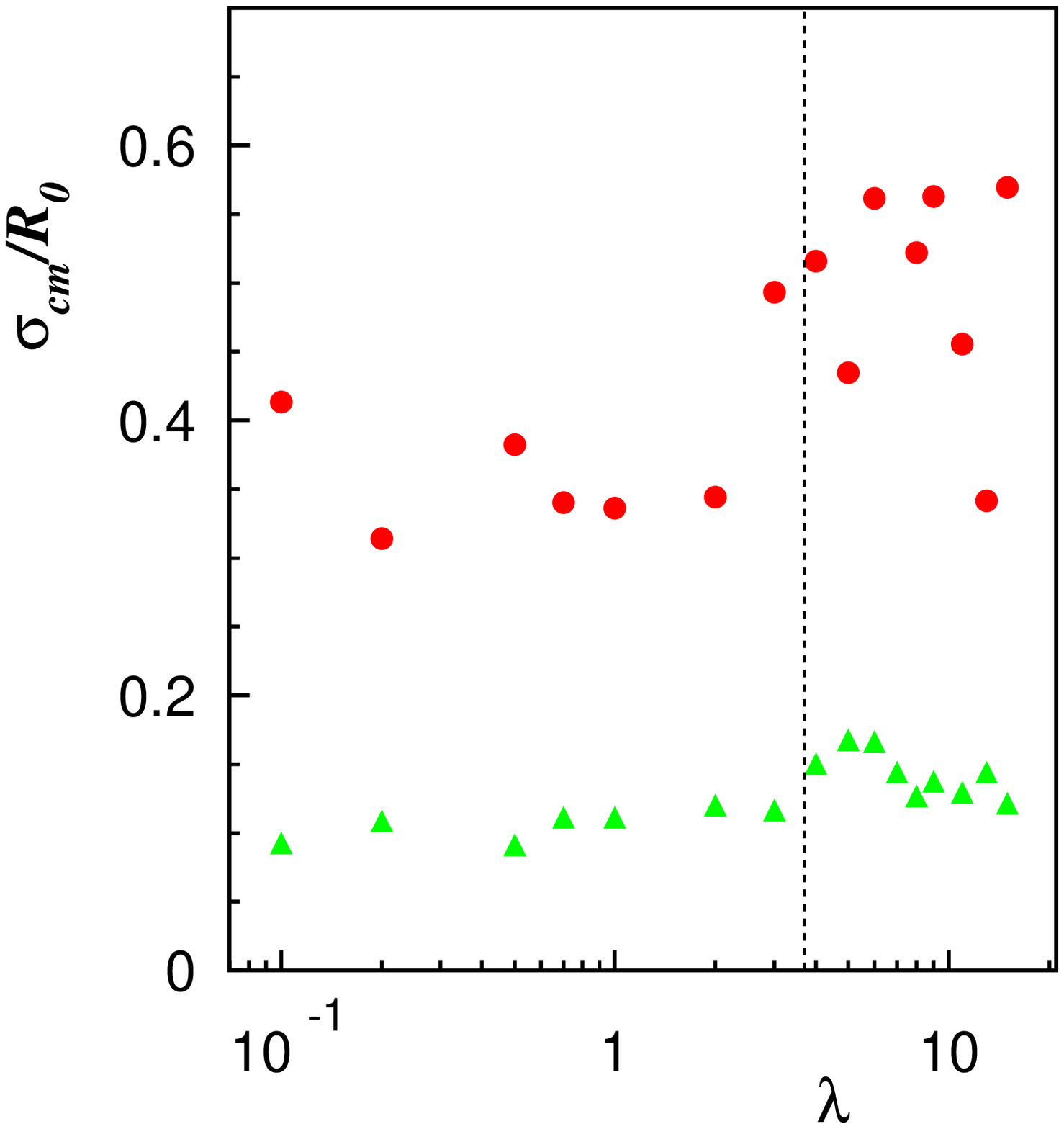}
\includegraphics*[width=.47\textwidth]{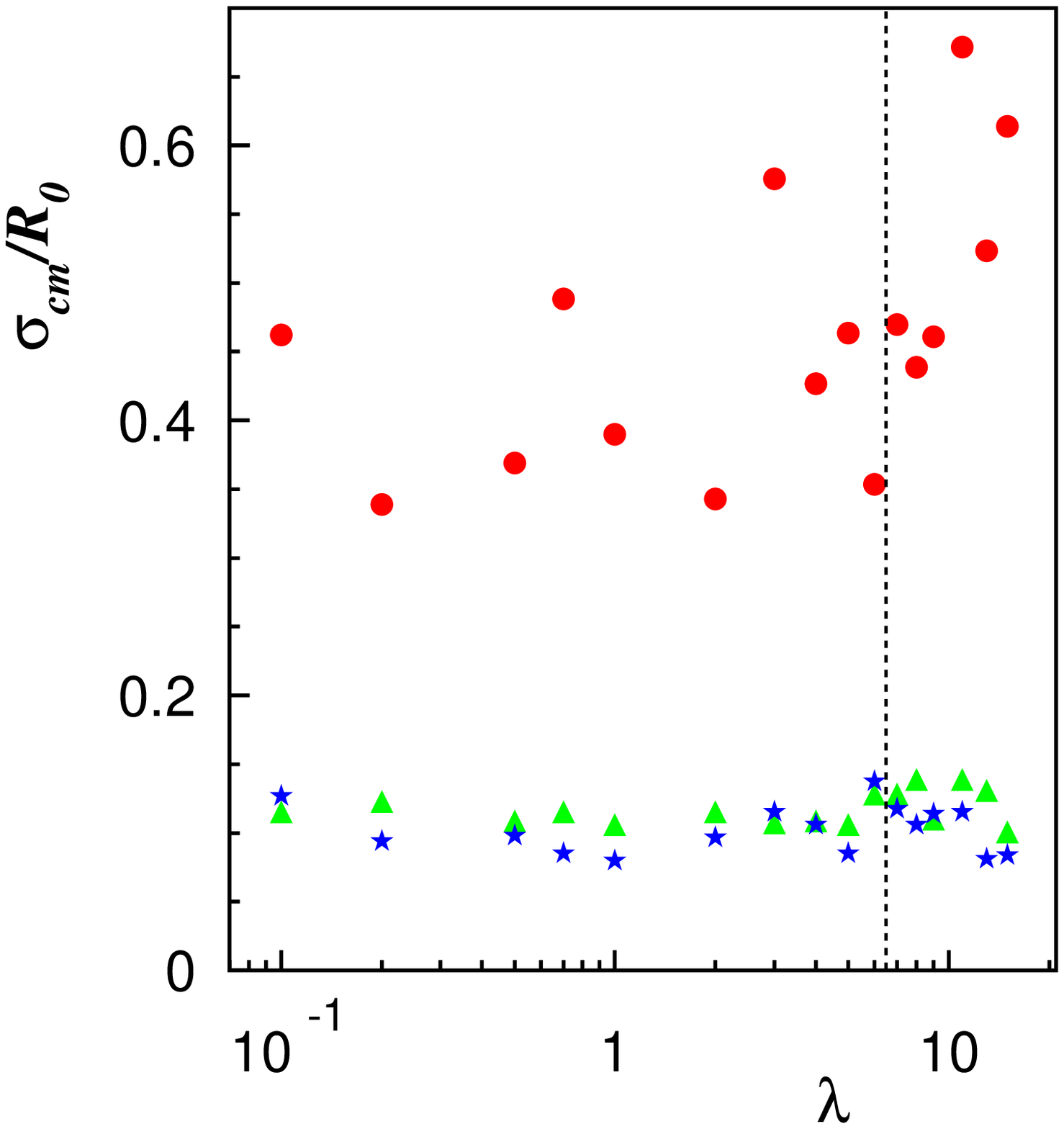}
\caption{Rms fluctuation values $\sigma_{cm}$ of the vertical position 
of the vesicle center of mass of Fig.~9
as a function of $\lambda$
at $S^{*}=0.80$ (left) and 
$S^{*}=0.95$ (right)
for $\kappa/(k_B T R_0)=6.58$ ($\bullet$), $65.8$ ($\blacktriangle$), $164.5$
($\star$). The tank-treading to tumbling transition
in the Keller-Skalak theory \cite{kell82} is marked by the dashed
vertical lines.\label{fig10}}
\end{figure}

The time behavior of vertical position $y_{cm}$ 
of the vesicle center of mass displays Brownian diffusion across the
channel width up to the longest simulated time, see Figure~\ref{fig9}.
The vesicle does not span the whole channel cross-section due to the lift
force which pushes it far from the walls \cite{mess09}.
In previous studies
\cite{thie13,kaou14}, 
where thermal noise is absent, vesicles move along the center line
of the channel without lateral displacement 
and with a regular arrangement in the TT
steady state, in two or three files at higher concentrations 
\cite{thie14,shen17}. It was later found that
there is a critical viscosity contrast above which the vesicle
can be either placed along the center line or off-centered without lateral
wandering \cite{nait18}.
The rms fluctuation values 
$\sigma_{cm}=\sqrt{\langle (\Delta y_{cm})^2 \rangle}$ 
are reported in Figure~\ref{fig10}. For the lowest values of the bending rigidity
it is evident that $\sigma_{cm}$ increases with the viscosity ratio 
$\lambda$ due to the more circular shape,
while this trend is less pronounced for further increasing
$\kappa/(k_B T R_0)$.
Moreover, a reduction in the values of $\sigma_{cm}$ can be appreciated when
increasing the bending rigidity with no significant dependence on the
reduced area $S^*$. In the TT regime it results to be
$\sigma_{cm}/R_0 \simeq \sqrt{(k_B T R_0)/\kappa} \propto  \sqrt{1/Pe}$ for the explored range
of bending rigidities. The term $\sqrt{(k_B T R_0)/\kappa}$ is the rms 
value of the vesicle deformation amplitude \cite{fink08}.

\begin{figure}[H]
\includegraphics*[width=.47\textwidth]{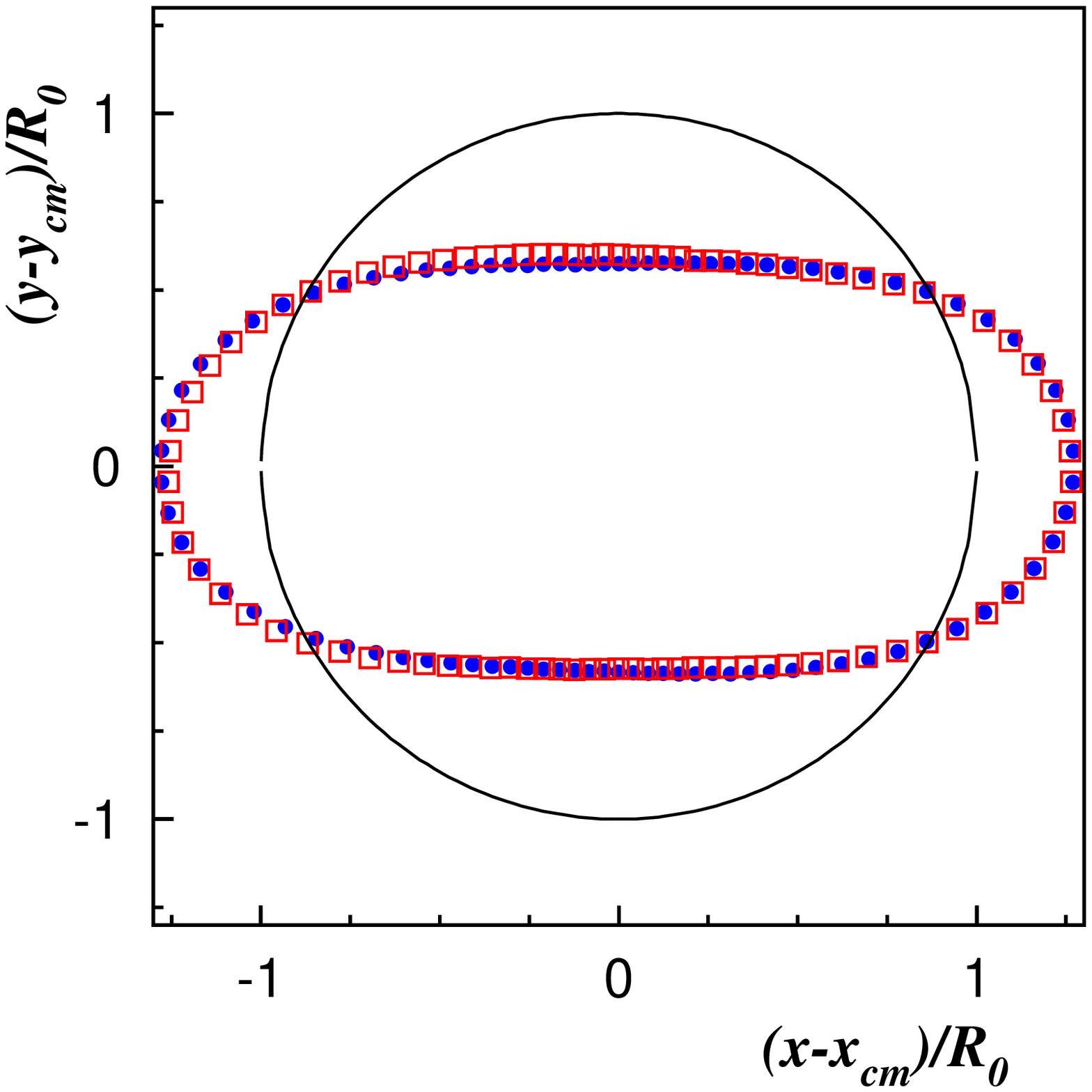}
\includegraphics*[width=.47\textwidth]{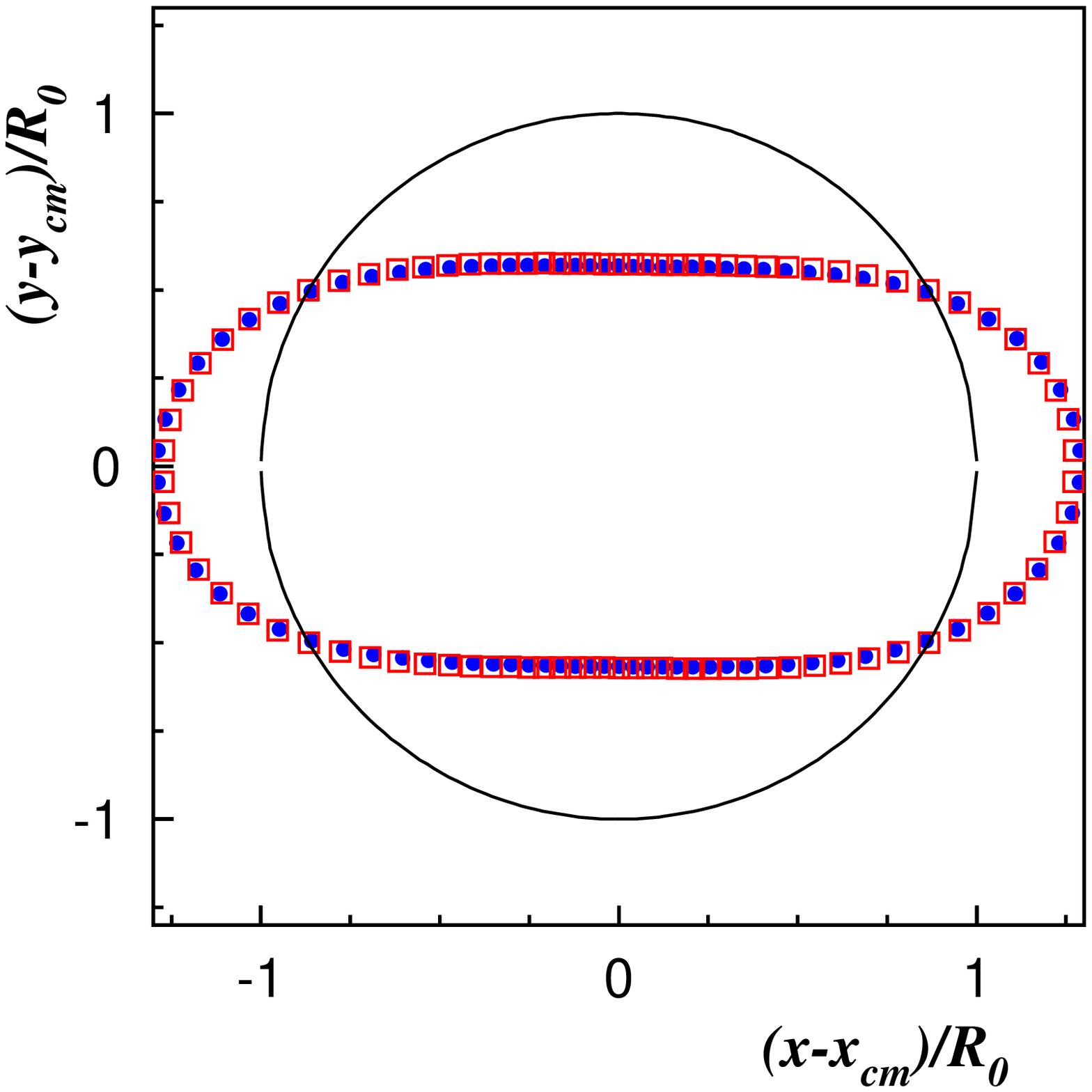}\\*
\includegraphics*[width=.47\textwidth]{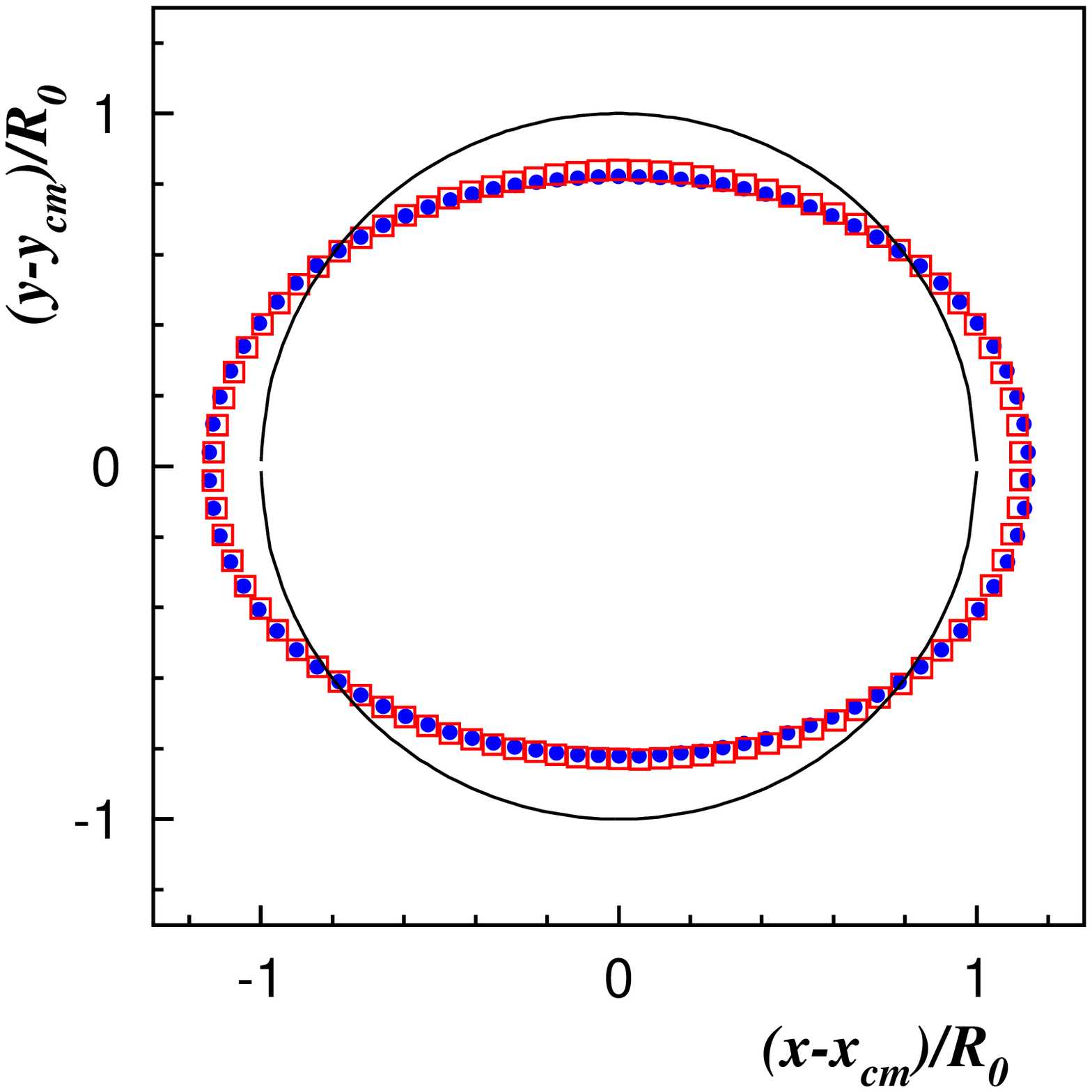}
\includegraphics*[width=.47\textwidth]{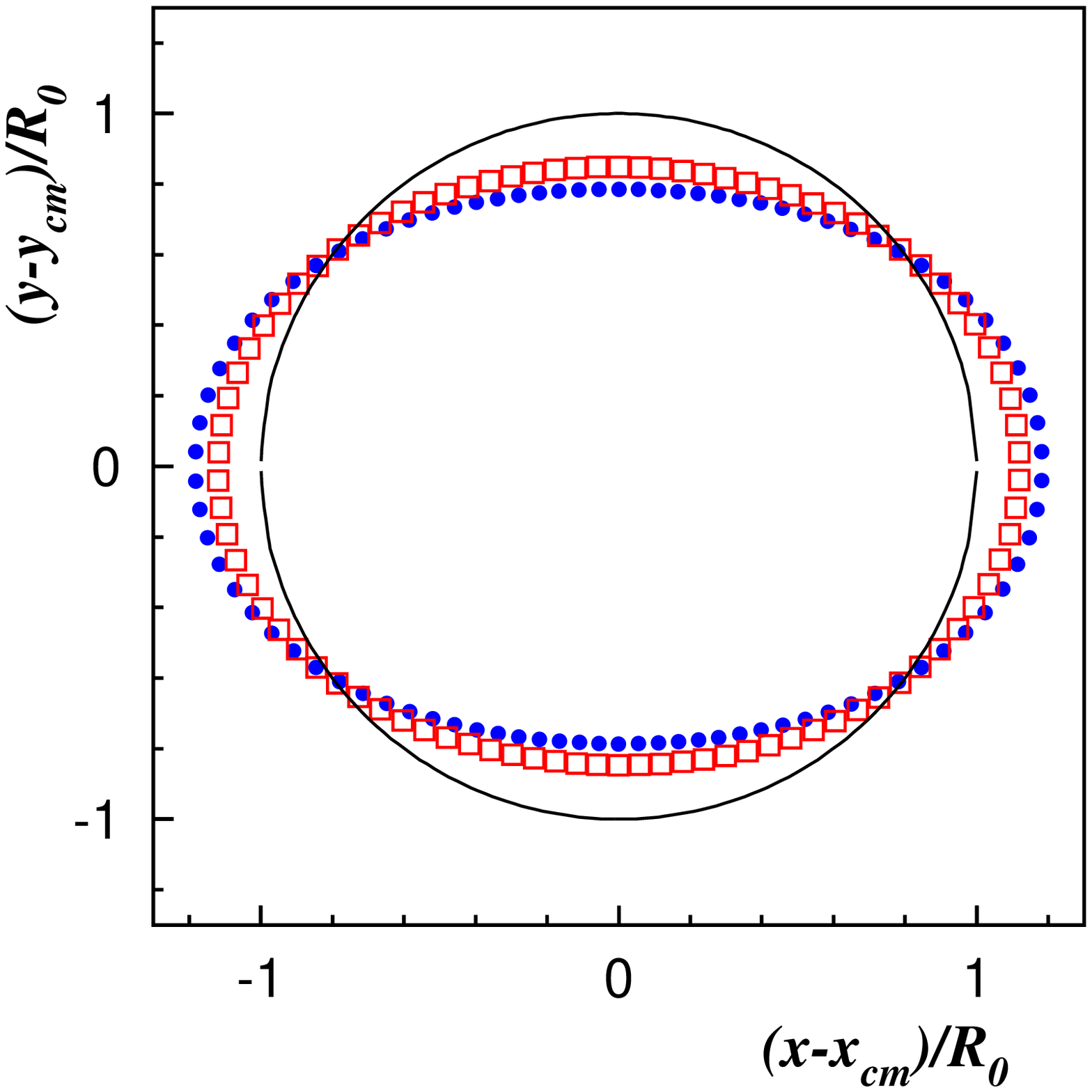}
\caption{Vesicle average configurations at
$S^{*}=0.80$ (upper row) and $S^{*}=0.95$ (lower row) 
for $\kappa/(k_B T R_0)=6.58$ (left), $65.8$ (right)
with $\lambda= 1 (\bullet), 11 (\Box)$. The full line represents the unit
circle as a reference.\label{fig11}}
\end{figure}

Finally, 
the average configurations of the vesicle 
are presented in Figure~\ref{fig11} for reduced area $S^*=0.80, 0.95$, 
bending rigidity  $\kappa/(k_B T R_0)=6.58, 65.8$, and 
two values of the viscosity contrast.
The shapes are obtained by averaging in time and space, in the vesicle
eigenvector reference frame, the positions of membrane beads in circular sectors
of width $\pi/45$ radians. 
This visualizes how 
the vesicle becomes more rounded going from the TT to the TU regime
in the case with  $\kappa/(k_B T R_0)=65.8$ at  $S^*=0.95$.
The reduction of the asphericity is less
appreciable in the other cases.

%%%%%%%%%%%%%%%%%%%%%%%%%%%%%%%%%%%%%%%%%%
\section{Discussion}

We can now relate
the observed behavior of the intrinsic viscosity
$\langle \eta_I \rangle$ in Fig.~2 to the
changes in vesicle shape and diffusion. 
We think that 
the monotonic growth of $\langle \eta_I \rangle$
is due to the interplay of several mechanisms.
As previously observed in Ref.~\cite{kant08},
shape ﬂuctuations favor energy dissipation that increases $\langle \eta_I \rangle$,
while alignment with the ﬂow direction causes a decrease of $\langle \eta_I \rangle$ with increasing
the viscosity contrast.
The vesicle becomes more rounded
with increasing $\lambda$ as 
indicated by the average asphericity. 
As a consequence 
the vesicle experiences 
a larger resistance to the flow with tilt angle approaching the
values $\pi/4$. This counteracts the reduction
due to the decrease of the average inclination angle 
when approaching the TT-to-TU transition. 
The most relevant effect due to thermal noise of the fluid is that the vesicle is not located 
at the center of the channel, but wanders across it 
due to fluctuation-induced Brownian diffusion 
(the possible influence of this
effect on the intrinsic viscosity
was already mentioned in Ref.~\cite{thie13}).
This implies that the vesicle can never move along the centerline of the channel,
which is the state of minimum dissipation when thermal effects are neglected \cite{thie13}.
The amplitude of this lateral motion is quantified
by $\sigma_{cm}$, which grows with increasing
viscosity ratio for the lowest value of the bending rigidity. 
Since the vesicle gets closer to the walls, 
a larger resistance of the vesicle to the flow might be induced,
similarly to what happens for colloids whose effective diffusion coefficient
reduces close to a wall \cite{bren61}.  
This effect would
contribute to the increase of $\langle \eta_I \rangle$ even in the TT regime.
We remark that since it results to be $\sigma_{cm} \propto \sqrt{1/Pe}$, as previously found,
much higher values of the Peclet number are required in order to access a regime
where $\sigma_{cm} \simeq 0$ to ignore thermal fluctuations. 

The outlined picture persists
with increasing the bending rigidity $\kappa$,
when
the value of $\langle \eta_I \rangle$ is reduced 
but its $\lambda$-dependence is not affected. Similar values 
of $\langle \eta_I \rangle$ are observed
for the highest bending rigidity
where the TT-to-TU 
transition is sharper and the vesicle
becomes more rigid, as observed in the values of the average
asphericity and of
the rms fluctuations
$\sigma_M$ and $\sigma_m$ which hardly change with $\lambda$.
In the TT regime the effect of increasing the bending rigidity is
to reduce
the average inclination angle $\langle \Theta \rangle$, its variance
$\sigma_{\Theta}$, and 
$\sigma_{cm}$ 
with respect to the case with the lowest bending rigidity, while
the vesicle appears to be less circular.
As a consequence the vesicle has less resistance to the flow, which 
explains the reduction of $\langle \eta_I \rangle$ 
when compared to lower values of $\kappa$.
In the TU regime, the difference in the average asphericity
for the three values of the bending rigidity diminishes, causing accordingly
a reduction in the difference of the average intrinsic viscosities.

To complete our discussion,  we note that 
it was argued in Ref.~\cite{thie13} that the monotonic behavior of 
$\langle \eta_I \rangle$
might be due to measurements done in short transient regimes 
but, as here shown, this is not the case. 
Moreover, 
our results do not depend on the choice either of the channel
length $L_x/R_0=19$
or of the degree of confinement $2R_0/L_y=0.35$, as suggested 
in Refs.~\cite{thie13,kaou14}.
Indeed these two values are intermediate between the ones used in those studies 
\cite{thie13,kaou14} where the
non-monotonic behavior of the intrinsic viscosity was observed without
thermal fluctuations. 
 
\section{Conclusions}

We believe that the
monotonic growth of $\langle \eta_I \rangle$ has to be related
to the presence of thermal fluctuations missing in other models.
This effect persists up to the highest Peclet number
of about $2 \times 10^3$. 
In a simplified stochastic three-dimensional model of vesicles
in shear flow \cite{nogu05} it was shown that thermal fluctuations
cannot be neglected  up to $Pe=1.2 \times 10^3$. Much higher values of $Pe$
are required, as previously discussed, in order to ignore
thermal fluctuations.
Finally, we add that the relevance of thermal noise in the
vesicle dynamics was demonstrated also for the VB regime in numerical
\cite{nogu07bis,mess09}, theoretical \cite{abre12,abre13}, 
and experimental studies \cite{kant06,leva12}.

%%%%%%%%%%%%%%%%%%%%%%%%%%%%%%%%%%%%%%%%%%
\vspace{6pt} 

%%%%%%%%%%%%%%%%%%%%%%%%%%%%%%%%%%%%%%%%%%
\funding{This research received no external funding.}

\dataavailability{Data are available upon reasonable request.} 

\acknowledgments{AL wishes to thank G. Gompper for useful discussions and
hospitality at Forschungszentrum J\"{u}lich.
This work was performed under the auspices of GNFM-INdAM.}

\conflictsofinterest{The author declares no conflict of interest.}

\appendixtitles{no} % Leave argument "no" if all appendix headings stay EMPTY (then no dot is printed after "Appendix A"). If the appendix sections contain a heading then change the argument to "yes".   
\appendixstart
\appendix
\section[\appendixname~\thesection]{}
Here the numerical implementation of the
algorithm is described for a system bounded by two moving solid walls.
In the presence of walls, the system consists of fluid
particles and virtual particles. The fluid particles represent the
solvent, while the virtual particles are required
to impose no-slip conditions at the walls.
First, the positions and velocities, for both the solvent particles
and the vesicle beads, are initialized . The fluid
particles are distributed uniformly inside the system with
average number $n$ of particles per cell.
An extra layer of collision cells is required next to the walls
to enforce boundary conditions.
The virtual particles are also uniformly distributed with
the same number density. All the fluid real
particles and beads are initialized
with velocities sampled from the Maxwell–Boltzmann
distribution with variances $k_B T/m$ and $k_B T/m_p$, respectively, and
zero mean. The velocities of virtual particles are from the Maxwell–Boltzmann
distribution with variances $k_B T/m$ and average
$\pm \frac{1}{2} \dot\gamma L_y$.
The initial linear and angular momentum are
removed from each cell and from all the beads, and
the velocities are rescaled to set the temperature to the value $T$.

At each time step Newton's equations of motion for beads are integrated
by means of the velocity-Verlet algorithm with
time step $\Delta t_p$ \cite{allen}.
Every $\Delta t_s/\Delta t_p$ time steps the MPC algorithm and the
solvent-vesicle collisions are performed in the following way:
\begin{enumerate}
\item All the solvent particles are streamed according to Equation
  (\ref{eq.prop}). Particles crossing walls undergo bounce-back collisions
  changing their velocities as
  ${\bf v}_i \rightarrow 2 {\bf v}_{wall} - {\bf v}_i$ where ${\bf v}_{wall}$
  and $- {\bf v}_{wall}$ are the wall velocities with
  ${\bf v}_{wall}=(v_{wall},0)$.
\item The solvent particles and the beads which overlap, are looked for
  and their velocities are modified according to Equation
  (\ref{eq.mix}).
\item Galilean invariance is violated when the mean-free path $l$ is much
  smaller than the cell size $a$.
To restore the Galilean invariance \cite{ihle2001}, all the fluid
particles are moved by a random
vector  ${\bf s}$ as
${\bf r}_i \rightarrow {\bf r}_i + {\bf s}$. The components of this
random vector are drawn from a uniform distribution
in the interval $[-a/2,a/2]$.
\item All solvent particles are sorted in respective cells and cell-level
  quantities are calculated.
\item The velocities ${\bf v}_i$ of fluid particles not scattering
  with the vesicle, are updated according to Equation (\ref{eq.coll}).
  The virtual particles are assigned a new random velocity.
  \item All fluid particles are shifted back to their original
position as ${\bf r}_i \rightarrow {\bf r}_i - {\bf s}$.
\end{enumerate}

%%%%%%%%%%%%%%%%%%%%%%%%%%%%%%%%%%%%%%%%%%

\begin{adjustwidth}{-\extralength}{0cm}
%\printendnotes[custom] % Un-comment to print a list of endnotes

\reftitle{References}

% Please provide either the correct journal abbreviation (e.g. according to the “List of Title Word Abbreviations” http://www.issn.org/services/online-services/access-to-the-ltwa/) or the full name of the journal.
% Citations and References in Supplementary files are permitted provided that they also appear in the reference list here. 

%=====================================
% References, variant A: external bibliography
%=====================================
%\bibliography{your_external_BibTeX_file}

%=====================================
% References, variant B: internal bibliography
%=====================================

\end{adjustwidth}
\end{document}